\documentclass[journal]{IEEEtran}

\usepackage{cite}
\ifCLASSINFOpdf
  \usepackage[pdftex]{graphicx}
\else
  \usepackage[dvips]{graphicx}
\fi

\usepackage[cmex10]{amsmath}
\usepackage{amsmath,amsfonts,amssymb}
\usepackage{mdwmath}
\usepackage{algorithm}
\usepackage{algorithmic}
\usepackage{array}
\usepackage{fixltx2e}
\usepackage{bm}
\usepackage{stfloats}
\usepackage{graphicx}
\usepackage{caption}
\usepackage{mdwtab}
\usepackage{subfig}
\usepackage{booktabs}
\usepackage{multirow}
\usepackage{etoolbox}
\usepackage{makecell}
\usepackage{xcolor}
\usepackage[bottom]{footmisc}
\usepackage{mdframed}

\newcommand{\e}{\begin{equation}}
\newcommand{\ee}{\end{equation}}
\newcommand{\eqn}{\begin{eqnarray}}
\newcommand{\eeqn}{\end{eqnarray}}

\begin{document}
\title{Closed-Loop Sparse Channel Estimation for Wideband Millimeter-Wave Full-Dimensional MIMO Systems}
\author{Anwen Liao,~\IEEEmembership{Student Member,~IEEE}, Zhen Gao,~\IEEEmembership{Member,~IEEE}, Hua Wang,\\
 Sheng Chen,~\IEEEmembership{Fellow,~IEEE}, Mohamed-Slim Alouini,~\IEEEmembership{Fellow,~IEEE}, and Hao Yin
\vspace*{-5.0mm}
\thanks{A. Liao, Z. Gao, and H. Wang are with School of Information and Electronics,
 Beijing Institute of Technology, Beijing 100081, China (E-mail: gaozhen16@bit.edu.cn).}
\thanks{S. Chen is with School of Electronics and Computer Science, University of
 Southampton, Southampton SO17 1BJ, UK, and also with King Abdulaziz University,
 Jeddah 21589, Saudi Arabia.}
\thanks{M.-S. Alouini is with the Electrical Engineering Program, Division of Physical
 Sciences and Engineering, King Abdullah University of Science and Technology, Thuwal,
 Makkah Province, Saudi Arabia.}
\thanks{H. Yin is with Institute of China Electronic System Engineering Corporation,
 Beijing 100141, China.}
}

\markboth{Journal of \LaTeX\ Class Files,~Vol.~x, No.~x, September~2019}%
{Liao \MakeLowercase{\textit{et al.}}: Manuscript of IEEEtran.cls for IEEE Journals}

\maketitle

\begin{abstract}
 This paper proposes a closed-loop sparse channel estimation (CE) scheme for wideband
 millimeter-wave hybrid full-dimensional multiple-input multiple-output and time
 division duplexing based systems, which exploits the channel sparsity in both angle
 and delay domains. At the downlink CE stage, random transmit precoding matrix is
 designed at base station (BS) for channel sounding, and receive combining matrices at user
 devices (UDs) are designed whereby the hybrid array is visualized as a low-dimensional digital
 array for facilitating the multi-dimensional unitary ESPRIT (MDU-ESPRIT) algorithm to
 estimate respective angle-of-arrivals (AoAs). At the uplink CE stage, the estimated
 downlink AoAs, namely, uplink angle-of-departures (AoDs), are exploited to design
 multi-beam transmit precoding matrices at UDs to enable BS to estimate the uplink AoAs,
 i.e., the downlink AoDs, and delays of different UDs, whereby the MDU-ESPRIT algorithm is used
 based on the designed receive combining matrix at BS. Furthermore, a maximum likelihood
 approach is proposed to pair the channel parameters acquired at the two stages, and
 the path gains are then obtained using least squares estimator. According to spectrum
 estimation theory, our solution can acquire the super-resolution estimations of the
 AoAs/AoDs and delays of sparse multipath components with low training overhead.
 Simulation results verify the better CE performance and lower computational complexity
 of our solution over state-of-the-art approaches.
\end{abstract}

\begin{IEEEkeywords}
 Wideband channel estimation, millimeter-wave, hybrid full-dimensional MIMO, super-resolution.
\end{IEEEkeywords}

\IEEEpeerreviewmaketitle

\section{Introduction}\label{S1}

 Millimeter-wave (mmWave) communication with the aid of massive multiple-input
 multiple-output (MIMO) is an enabling technology for next-generation mobile
 communications, since the abundant spectrum resources at mmWave frequency band can
 boost the throughput by orders of magnitude \cite{TWC_XiaoZY16,Gao_UDN15}. To
 mitigate the severe path loss for mmWave signal, massive MIMO is usually integrated
 into mmWave communications to form beams for directional signal transmission
 \cite{TSP_Mawenyan18,JSAC_HuangYM18}. However, the powerful fully-digital MIMO
 architecture, which requires a radio frequency (RF) chain for each antenna, is
 unaffordable for mmWave massive MIMO, due to the prohibitive hardware cost and power
 consumption of RF chains required \cite{TSP_LiuAn16}. The hybrid MIMO architecture
 with a much smaller number of RF chains than that of antennas offers a practical
 solution by using hybrid analog/digital beamforming \cite{JSTSP_LiuAn18}.
 Nonetheless, for such a hybrid MIMO system, it is challenging to estimate the
 high-dimensional mmWave channel from the low-dimensional effective measurements
 observed from the limited number of RF chains, since the training overhead for
 channel estimation (CE) can be excessively high \cite{Gao_UDN15}. Moreover, the low
 signal-to-noise ratio (SNR) before beamforming can further degrade the performance
 of channel state information (CSI) acquisition \cite{Tcom_HuangYM17}.

\subsection{Related Work}\label{S1.1}

 Several approaches were proposed in the literature to acquire CSI for narrowband
 mmWave communications, including codebook-based beam training
\cite{IEEE802.11ad,TSP_Overlapped_Beam17,TWC_XiaoZY17,Access_XiaoZY17} and compressed
 sensing (CS)-based CE \cite{Tcom_OMP16,Gu_CL18}. The beam training
 approaches were initially adopted in analog beamforming, e.g., IEEE standards
 802.11ad \cite{IEEE802.11ad} and 802.15.3c \cite{Gao_UDN15}, where the transceiver
 exhaustively searches for the optimal beam pair from a predefined codebook to
 maximize the received SNR for improved transmission performance. To reduce the search
 dimension of codebooks for achieving lower training overhead, the multi-stage
 overlapped beam patterns were designed in \cite{TSP_Overlapped_Beam17}, where the
 beam patterns can become narrow as the training stage increases. However, these
 schemes only consider the analog beamforming with single-stream transmission. For
 hybrid beamforming with multi-stream transmission, beam training solutions with
 hierarchical multi-beam codebooks were proposed in \cite{TWC_XiaoZY17,Access_XiaoZY17},
 where the optimal multi-beam pairs can be acquired after hierarchical beam search
 with gradually finer and narrower beams. However, the training overhead of a beam
 training scheme is usually proportional to the dimension of codebook, which is
 very large for full-dimensional (FD) MIMO with a large number of antennas. By
 exploiting the inherent angle-domain sparsity of mmWave MIMO channels, several
 CS-based CE schemes were proposed to reduce the CE overhead \cite{Tcom_OMP16,Gu_CL18}.
 In \cite{{Tcom_OMP16}}, the orthogonal matching pursuit (OMP) algorithm was considered
 to estimate sparse mmWave channels by formulating the CSI acquisition problem as a
 sparse signal recovery problem, where a redundant dictionary with non-uniformly
 quantized angle-domain girds was designed for improved performance. Furthermore,
 a Bayesian CS-based CE scheme was proposed in \cite{Gu_CL18} by considering the
 impact of transceiver hardware impairments. Besides, by leveraging the low-rank
 property of mmWave channels, a CANDECOMP/PARAFAC decomposition-based CE scheme
 \cite{TWC_Zhou16} was proposed with further improved performance.

 The aforementioned solutions \cite{TSP_Overlapped_Beam17,TWC_XiaoZY17,Access_XiaoZY17,Tcom_OMP16,Gu_CL18,TWC_Zhou16}
 only consider frequency-flat mmWave channels but practical mmWave channels can be
 frequency selective due to the very large system bandwidth in mmWave frequency band
 and the distinct delay spreads of multipath components (MPCs) \cite{JSTSP_Heath16}.
 A distributed grid matching pursuit (DGMP) algorithm was proposed in \cite{Gao_CL16}
 to estimate time-dispersive channels, where orthogonal frequency division multiplexing
 (OFDM) is considered. An adaptive grid matching pursuit (AGMP) algorithm developed
 from the DGMP was proposed to reduce power leakage by using adaptive grid matching solution
 \cite{VTC_Spring17}. In \cite{JSAC_OMP17}, the sparse mmWave channels at different
 subcarriers were estimated separately by utilizing the OMP, but the computational
 complexity is high as the number of subcarriers is typically large. To reduce complexity,
 a simultaneous weighted (SW)-OMP based scheme was proposed in \cite{TWC_SW_OMP18},
 which  exploits the angle-domain common sparsity of channels at different subcarriers
 to improve performance. By leveraging the common sparsity of delay-domain channels
 among transceiver antenna pairs, a block CS-based CE solution was proposed for
 mmWave fully-digital MIMO system \cite{JSAC_YangF17}, where the training sequences
 are designed to improve CE performance. Based on the low-rank property of wideband
 mmWave channels, the training signal received can be formulated as a high-order
 tensor with the low-rank CANDECOMP/PARAFAC decomposition to estimate the dominated
 channel parameters, including angle-of-arrivals/angle-of-departures (AoAs/AoDs)
 and delays \cite{JSAC_Zhou17}. However, most CS-based CE schemes for wideband
 mmWave MIMO usually adopt discrete AoAs/AoDs grids in CS dictionary, but the practical
 AoAs/AoDs of MPCs are continuously distributed. This mismatch
 may degrade CE performance. Moreover, the state-of-the-art works
\cite{TSP_Overlapped_Beam17,TWC_XiaoZY17,Access_XiaoZY17,Tcom_OMP16,Gu_CL18,TWC_Zhou16,Gao_CL16,VTC_Spring17,JSAC_YangF17,JSAC_OMP17,TWC_SW_OMP18,JSAC_Zhou17,TSP_Gao_FF18}
 usually focus on the ideal uniform linear array (ULA) while seldom investigate the
 practical uniform planar array (UPA). Compared to the ULA, the UPA offers more
 compact array with three-dimensional (3-D) beamforming in both horizontal and vertical
 directions \cite{Tcom_ZhengLe18,TVT_HuChen18}, leading to the FD-MIMO. Although
 mmWave FD-MIMO CE has been investigated in \cite{Tcom_ZhengLe18} and \cite{TVT_HuChen18},
 they only considered either fully-digital MIMO or frequency-flat channels.

\begin{figure}[!tp]
\begin{center}
\includegraphics[width=0.7\columnwidth, keepaspectratio]{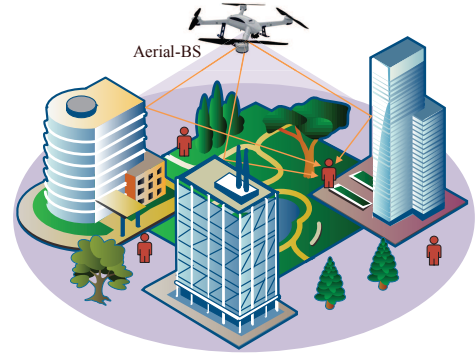}
\end{center}
\captionsetup{font = {footnotesize}, name = {Fig.}, labelsep = period}\
\caption{The air-ground mmWave channels between the UAV aerial-BS and UDs exhibit
 sparsity in both angle-domain and delay-domain due to the limited significant
 scatterers \cite{CM_XiaoZY16,WC_UAV19}.}
\label{FIG1}
\end{figure}

\begin{figure*}[!tp]
\begin{center}
\includegraphics[width = 1.6\columnwidth, keepaspectratio]{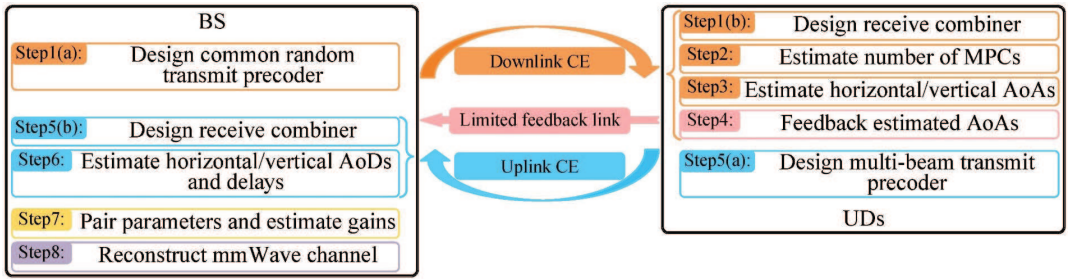}
\end{center}
\captionsetup{font = {footnotesize}, singlelinecheck = off, justification = raggedright, name = {Fig.}, labelsep = period}%
\caption{Procedure of the proposed closed-loop sparse CE solution, where the limited
 feedback is realized via the low-frequency control link.}
\label{FIG2}
\end{figure*}

\begin{figure*}[!tp]
\begin{center}
\includegraphics[width = 1.75\columnwidth,keepaspectratio]{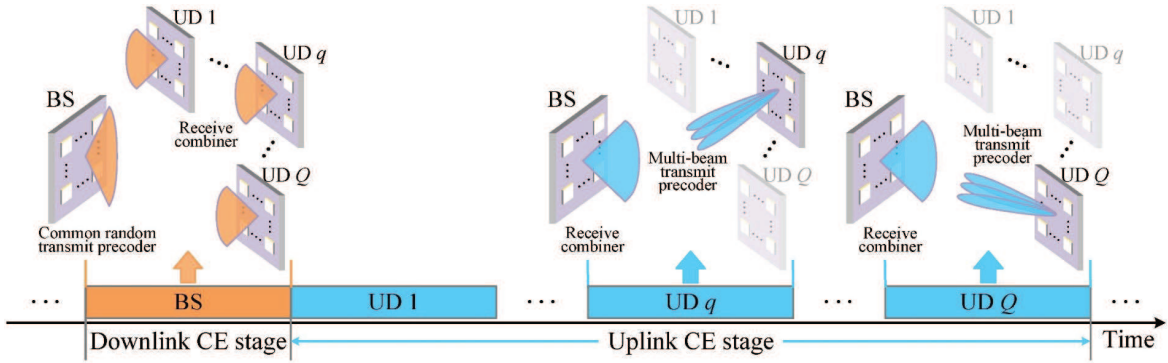}
\end{center}
\captionsetup{font = {footnotesize}, singlelinecheck = off, justification = raggedright, name = {Fig.}, labelsep = period}%
\caption{Frame structure of the proposed closed-loop sparse CE solution.}
\label{FIG3}
\end{figure*}

\subsection{Our Contributions}\label{S1.2}

 We propose a closed-loop sparse CE scheme for multi-user wideband mmWave FD-MIMO
 systems by exploiting the sparsity of MPCs in both angle and delay domains. To
 illustrate this sparsity, we consider the mmWave FD-MIMO based unmanned aerial vehicle
 (UAV) aerial-base station (BS), as shown in Fig.~\ref{FIG1}, which has flexible deployment
 capacity to serve user devices (UDs) in hotspot areas \cite{WC_AerialBS18}. Different from
 the terrestrial BS in 3GPP or QuaDRiGa \cite{Tech_Rep_QuaDRiGa}, \cite{TAP_QuaDRiGa}, UAV
 aerial-BS usually works at the height of hundreds of meters, where fewer MPCs corresponding
 to dominant scatterers could establish the communications links between the aerial-BS and UD.
 Therefore, the air-ground mmWave channels in such aerial-BS based systems exhibit inherent
 sparsity in both angle and delay domains due to the limited significant scatterers. By
 carefully designing the transmit beamforming or precoding and receive combining in the training
 stage, our solution is capable of acquiring the super-resolution estimates of AoAs, AoDs,
 and delays\footnote{By contrast, the state-of-the-art CS-based solutions \cite{Gao_CL16,VTC_Spring17,JSAC_OMP17,TWC_SW_OMP18}
 only focus on the angle-domain sparsity and design the limited resolution CS dictionary
 with quantized angle-domain grids. This quantization error limits the achievable CE performance.}
 based on spectrum estimation theory \cite{2D_U_ESPRIT96,TSP_JADE98} with low training
 overhead and computational complexity. In terms of sparse mmWave UAV air-ground channels,
 the proposed CE scheme can obtain a better performance of parametric CE. To clearly show
 the novelty and new contribution of our proposed solution as well as to contrast it with
 the existing solutions, below we conceptually explain our proposed closed-loop sparse CE scheme.

 Our closed-loop solution includes the downlink CE stage followed by the uplink
 CE stage as illustrated in Fig.~\ref{FIG2}, where the channel reciprocity in time
 division duplex (TDD) based systems is exploited \cite{JSAC_Yuan17,WC_GaoZ18,Tcom_ZhangS19}.
 In TDD based systems, downlink AoAs (AoDs) are uplink AoDs (AoAs). The frame
 structure of our solution is further depicted in Fig.~\ref{FIG3}. As shown in
 Figs.~\ref{FIG2} and \ref{FIG3}, at the downlink CE stage, the horizontal/vertical
 AoAs of sparse MPCs are first estimated at each UD and they are fed back to the
 BS with limited quantization accuracy through the feedback link. At this stage,
 we design a common random transmit precoding matrix at the BS to transmit the
 training signals for omnidirectional channel sounding and we design the
 receive combining matrix at each UD to visualize the high-dimensional hybrid array as a
 low-dimensional digital array, which facilitates the use of the multi-dimensional
 unitary ESPRIT (MDU-ESPRIT) algorithm to estimate channel parameters. Similarly,
 at the uplink CE stage, the horizontal/vertical AoDs and delays associated with
 different UDs are successively estimated by using the MDU-ESPRIT algorithm at
 the BS. Owing to the channel reciprocity, the AoAs estimated at UD side can be
 utilized as \emph{a priori} to design the multi-beam transmit precoding matrix to
 improve the received SNR for uplink CE. A maximum likelihood (ML) approach is
 adopted at the BS to pair the channel parameters acquired at these two stages and,
 consequently, the associated path gains can readily be obtained using the least
 squares (LS) estimator. Finally, the mmWave channel associated with each UD can
 be separately reconstructed based on the dominant channel parameters estimated
 above.

 In contrast to the existing solutions, our main contributions are summarized as
 follows:
\begin{itemize}
\item \textbf{We propose a closed-loop sparse CE solution with the common
 downlink CE stage for all UDs and the dedicated uplink CE stage for each UD}.
 At the downlink CE stage, the BS fully exploits its large transmit power to
 allow multiple UDs simultaneously perform CE for reduced training overhead,
 and only horizontal/vertical AoAs estimation with low computational complexity
 is required at UDs. At the uplink CE stage, the multi-beam transmit precoding
 matrices are designed at UDs to enhance CE accuracy, and the BS with high
 computational capacity can jointly estimate horizontal/vertical AoDs and delays.
 By contrast, the state-of-the-art CS-based solutions \cite{Gao_CL16,VTC_Spring17,JSAC_OMP17,TWC_SW_OMP18}
 are all based on open-loop approach, which imposes high computational complexity
 and storage requirements on the receiver\footnote{More specifically, to improve
 the CE accuracy, CS-based solutions \cite{Gao_CL16,VTC_Spring17,JSAC_OMP17,TWC_SW_OMP18}
 adopt the redundant dictionary, whose dimension is very high for mmWave massive
 MIMO. Therefore, downlink open-loop CE imposes prohibitive storage of redundant
 dictionary and the associated computational complexity on UDs, while uplink
 open-loop CE suffers from the low received SNR and thus poor CE performance due to
 the limited transmit power of UDs.}. It is worth emphasizing that by exploiting
 the horizontal/vertical AoAs estimated at the downlink CE stage, the multi-beam
 transmit precoding matrices designed at UDs significantly improve received SNR,
 which works even for the case that the number of MPCs is larger than that of RF
 chains, i.e., the number of beams can be larger than that of RF chains.
\item \textbf{We design the receive combining matrices at UDs and BS for visualizing the
 hybrid array as a digital array, to enable the application of spectrum
 estimation techniques}. For hybrid MIMO, it is challenging to directly apply
 spectrum estimation techniques to estimate channel parameters, since the
 shift-invariance structure of array response matrix observed from the digital
 baseband domain does not hold \cite{TSP_JADE98,TSP_Vandermonde98}. Our solution
 sheds light on how to apply spectrum estimation techniques, e.g., ESPRIT-type
 algorithms, to hybrid MIMO, so that the super-resolution estimation of channel
 parameters can be acquired with low training overhead and computational complexity.
 By contrast, to achieve the high-resolution estimations of channel parameters,
 the existing CS-based solutions \cite{Gao_CL16,VTC_Spring17,JSAC_OMP17,TWC_SW_OMP18}
 usually rely on the high-dimensional angle-domain or delay-domain redundant
 dictionary, which poses the excessively high computational complexity and
 storage requirements for FD-MIMO systems with large-scale antenna arrays.
\item \textbf{The double sparsity of MPCs in both angle and delay domains is
 harnessed in our proposed CE scheme}. By leveraging the double sparsity, our
 scheme formulates the CE problem as a multi-dimensional spectrum estimation
 problem, where the super-resolution estimations of horizontal/vertical
 AoAs/AoDs and delays can be obtained simultaneously. By comparison, the
 existing CE solutions \cite{Gao_CL16,VTC_Spring17,TWC_SW_OMP18} only
 consider the angle-domain sparsity of mmWave channels. Moreover, the
 delay-domain CE approaches of \cite{JSAC_OMP17,JSAC_YangF17} have to estimate
 the effective delay-domain channel impulse response (CIR), which includes the
 time-domain pulse shaping filter (PSF) that can weaken the delay-domain
 channel sparsity. By contrast, the super-resolution estimation of delays in
 our solution is immune to PSF.
\end{itemize}

 Throughout this paper, boldface lower and upper-case symbols denote column
 vectors and matrices, respectively. $(\cdot )^*$, $(\cdot )^{\rm T}$,
 $(\cdot )^{\rm H}$, $(\cdot )^{-1}$, $\left\lceil\cdot \right\rceil$ and
 $\left\lfloor\cdot\right\rfloor$ denote the conjugate, transpose, Hermitian
 transpose, matrix inversion, integer ceiling and integer floor operators,
 respectively. $\|\bm{a}\|_1$ and $\|\bm{a}\|_2$ are ${\ell_1}$-norm and
 ${\ell_2}$-norm of $\bm{a}$, respectively, while $\|\bm{A}\|_F$ is Frobenius
 norm of $\bm{A}$, and $|{\cal Q}|_c$ is the cardinality of the set ${\cal Q}$.
 The Kronecker and Khatri-Rao product operations are denoted by $\otimes$ and
 $\odot$, respectively. $\bm{I}_n$ denotes the $n\times n$ identity matrix and
 $\bm{O}_{m\times n}$ is the null matrix of size $m\times n$, while $\bm{1}_n$
 ($\bm{0}_n$) denotes the vector of size $n$ with all the elements being $1$
 ($0$). $\text{diag}(\bm{a})$ is the diagonal matrix with the elements of
 $\bm{a}$ at its diagonal entries, $\text{vdiag}(\bm{A})$ denotes the vector
 consisting of the main diagonal elements of $\bm{A}$, and $\text{Bdiag}([\bm{A}_1
 \cdots \bm{A}_n])$ denotes the block diagonal matrix with $\bm{A}_1,\cdots ,
 \bm{A}_n$ as its block diagonal entries. The expectation and determinant
 operators are denoted by $\mathbb{E}(\cdot )$ and $\det(\cdot )$, respectively.
 The modulo operation $\text{mod}(m,n)$ returns the remainder of dividing $m$
 by $n$, and $\text{mod}({\cal Q},n)$ returns the set containing $\text{mod}(m,n)$
 $\forall m\in {\cal Q}$ of the ordered set ${\cal Q}$. The operator
 $\text{find}(\bm{a}\neq 0)$ returns the set containing the indices of nonzero
 elements of $\bm{a}$, and $\text{mat}(\bm{a};m,n)$ converts the vector $\bm{a}$
 of size $m n$ into the matrix of size $m\times n$ by successively selecting every
 $m$ elements of $\bm{a}$ as its columns. The operator $\text{vec}(\bm{A})$
 stacks the columns of ${\bm{A}}$ on top of each another, $[\bm{A}]_{m,n}$ denotes
 the $m$th-row and $n$th-column element of $\bm{A}$, and $\bm{a}_{[m:n]}$ is the
 vector consisting the $m$th to $n$th elements of $\bm{a}$, while $\bm{A}_{[:,m:n]}$
 is the sub-matrix containing the $m$th to $n$th columns of $\bm{A}$.
 $\bm{A}_{[{\cal Q},:]}$ denotes the sub-matrix containing the rows of $\bm{A}$
 indexed in the ordered set ${\cal Q}$, and $\bm{A}_{[{\cal Q},i]}$ is the $i$th
 column of $\bm{A}_{[{\cal Q},:]}$. Finally, $\Re \{\cdot\}$ and $\Im \{\cdot\}$
 denote the real part and imaginary part of the argument, respectively.

\section{Downlink Channel Estimation Stage}\label{S2}

 Consider the mmWave FD-MIMO system with hybrid beamforming, where the BS and $Q$
 UDs are all equipped with UPA, and OFDM with $K$ subcarriers is adopted, while
 $N_{\rm s}^d$ independent signal streams are transmitted on each subcarrier \cite{Gao_CL16}.
 The BS (UD) employs $N_{\rm BS}=N_{\rm BS}^{\rm h} N_{\rm BS}^{\rm v}$
 ($N_{\rm UD}=N_{\rm UD}^{\rm h} N_{\rm UD}^{\rm v}$) antennas and
 $N_{\rm BS}^{\rm RF} \ll N_{\rm BS}$ ($N_{\rm UD}^{\rm RF} \ll N_{\rm UD}$)
 RF chains, where $N_{\rm BS}^{\rm h}$ ($N_{\rm UD}^{\rm h}$) and $N_{\rm BS}^{\rm v}$
 ($N_{\rm UD}^{\rm v}$) are the numbers of antennas in horizontal and vertical
 directions at the BS (UD), respectively.

\subsection{Downlink Channel Estimation Signal Model}\label{S2.1}

 The downlink CE stage lasts $N_d$ time slots and each time slot contains
 $N_{\rm o}^d$ OFDM symbols. The signal $\bm{y}_q[k,i,m]\! \in\!
 \mathbb{C}^{N_{\rm s}^d}$ received by the $q$th UD at the $k$th subcarrier of the
 $i$th OFDM symbol in the $m$th time slot can be expressed as
\begin{equation}\label{eq_y} 
\begin{split}
 \bm{y}_q[k,i,m] =& \bm{W}_{d,q}^{\rm H}[k,m] \bm{H}_q[k] \bm{F}_{d}[k,m] \bm{s}[k,i,m]\\
 \ & + \bm{W}_{d,q}^{\rm H}[k,m]  \bm{n}_q[k,i,m] ,
\end{split}
\end{equation}
 for $1\! \le\! q\! \le\! Q$, $0\! \le\! k\! \le\! K-1$, $1\! \le\! i\! \le\!
 N_{\rm o}^d$ and $1\! \le\! m\! \le\! N_d$. In (\ref{eq_y}), the UD's receive combining
 matrix $\bm{W}_{d,q}[k,m]\! =\! \bm{W}_{{\rm RF},q}[m] \bm{W}_{{\rm BB},q}[k,m]\! \in\!
 \mathbb{C}^{N_{\rm UD}\times N_{\rm s}^d}$ in which $\bm{W}_{{\rm RF},q}[m]\!
 \in\! \mathbb{C}^{N_{\rm UD}\times N_{\rm UD}^{\rm RF}}$ and
 $\bm{W}_{{\rm BB},q}[k,m]\! \in\! \mathbb{C}^{N_{\rm UD}^{\rm RF}\times N_{\rm s}^d}$
 are the analog and digital receive combining matrices, and the BS's transmit
 precoding matrix $\bm{F}_d[k,m]\! =\! \bm{F}_{{\rm RF},d}[m]\bm{F}_{{\rm BB},d}[k,m]\! \in\!
 \mathbb{C}^{N_{\rm BS}\times N_{\rm s}^d}$ in which $\bm{F}_{{\rm RF},d}[m]\! \in\!
 \mathbb{C}^{N_{\rm BS}\times N_{\rm BS}^{\rm RF}}$ and $\bm{F}_{{\rm BB},d}[k,m]\!
 \in\! \mathbb{C}^{N_{\rm BS}^{\rm RF}\times N_{\rm s}^d}$ are the analog and digital
 transmit precoding matrices, respectively, while $\bm{H}_q[k]\! \in\!
 \mathbb{C}^{N_{\rm UD}\times N_{\rm BS}}$ is the corresponding downlink channel matrix,
 $\bm{s}[k,i,m]\! \in\! \mathbb{C}^{N_{\rm s}^d}$ is the training signal with
 $\mathbb{E}\left(\bm{s}[k,i,m]\bm{s}^{\rm H}[k,i,m]\right)\! =\! \frac{1}{N_{\rm s}^d}
 \bm{I}_{N_{\rm s}^d}$, and $\bm{n}_q[k,i,m]\! \in\! \mathbb{C}^{N_{\rm UD}}$ is the
 complex additive white Gaussian noise (AWGN) vector with the covariance matrix
 $\sigma_n^2\bm{I}_{N_{\rm UD}}$, i.e., $\bm{n}_q[k,i,m]\! \sim\! {\cal CN}\!
 \left(\bm{0}_{N_{\rm UD}},\sigma _n^2 \bm{I}_{N_{\rm UD}}\right)$. Due to the
 constant modulus of the phase shift network (PSN), $\left[\bm{F}_{{\rm RF},d}[m]
 \right]_{j_1,j_2}\! =\! \frac{1}{\sqrt{N_{\rm BS}}}e^{\textsf{j}\vartheta_{1,j_1,j_2}}$
 and $\left[\bm{W}_{{\rm RF},q}[m]\right]_{j_1,j_2}\! =\! \frac{1}{\sqrt{N_{\rm UD}}}
 e^{\textsf{j}\vartheta_{2,j_1,j_2}}$ with $\vartheta_{1,j_1,j_2},\vartheta_{2,j_1,j_2}\!
 \in\! {\cal A}$, and ${\cal A}$ is the quantized phase set of the PSN with the
 resolution $N_q^{\rm ps}$, given by
\begin{align}\label{Set_A} 
 {\cal A} =& \left\{ -\pi ,-\pi +{\textstyle{2\pi \over 2^{N_q^{\rm ps}}}},-\pi +2 \cdot {\textstyle{2\pi \over 2^{N_q^{\rm ps}}}},
  \cdots, \pi - {\textstyle{2\pi \over 2^{N_q^{\rm ps}}}} \right\} .
\end{align}
 Also $\left\|\bm{F}_d[k,m]\right\|_F^2 \le N_{\rm BS}^{\rm RF}$ to guarantee the
 constraint on the total transmit power \cite{TSP_LiuAn16}. Here at the CE stage, some
 elegant solutions \cite{TSP_CSI_TDD06,XiaoZY_TWC13} can be used to acquire the robust
 synchronization of burst training signals without the knowledge of noise/interference
 power even at low SNR.

 Due to the obviously resolvable delay spread for each MPC caused by the large bandwidth,
 according to the typical mmWave channel model \cite{JSAC_Zhou17,Gao_CL16,VTC_Spring17,JSAC_OMP17,TWC_SW_OMP18,JSAC_YangF17,CM_XiaoZY16}, the downlink
 delay-domain continuous channel matrix $\bm{H}_q(\tau)\in \mathbb{C}^{N_{\rm UD}\times N_{\rm BS}}$
 with $L_q$ MPCs can be expressed as
\begin{align}\label{delay_domain_channel} 
 \bm{H}_q(\tau ) =& \beta_q \sum\limits_{l=1}^{L_q} \bm{H}_{q,l} p\left(\tau - \tau_{q,l}\right) ,
\end{align}
 where $\beta_q=\sqrt{N_{\rm UD}N_{\rm BS}/L_q}$ is the normalization factor, $\tau_{q,l}$
 is the delay of the $l$th MPC, and $p(\tau )$ denotes the equivalent PSF, while the
 complex gain matrix $\bm{H}_{q,l}\in \mathbb{C}^{N_{\rm UD}\times N_{\rm BS}}$ is given by
\begin{equation}\label{channel} 
 \bm{H}_{q,l} = \alpha_{q,l} \bm{a}_{\rm UD}\left( \mu_{q,l}^{\rm UD},\nu_{q,l}^{\rm UD}\right)
  \bm{a}_{\rm BS}^{\rm H}\left( \mu_{q,l}^{\rm BS},\nu_{q,l}^{\rm BS}\right) ,
\end{equation}
 where $\alpha_{q,l}\! \sim\! {\cal CN}(0,\sigma_\alpha^2)$ is the associated complex
 path gain, $\mu_{q,l}^{\rm UD}\! =\! \pi\sin\big(\theta_{q,l}^{\rm UD}\big)
 \cos\big(\varphi_{q,l}^{\rm UD}\big)$ ($\mu_{q,l}^{\rm BS}\! =\! \pi
 \sin\big(\theta_{q,l}^{\rm BS}\big)\cos\big(\varphi_{q,l}^{\rm BS}\big)$) and
 $\nu_{q,l}^{\rm UD}\! =\! \pi\sin\big(\varphi_{q,l}^{\rm UD}\big)$ ($\nu_{q,l}^{\rm BS}
 \! =\! \pi\sin\big(\varphi_{q,l}^{\rm BS}\big)$) denote the horizontally and vertically
 spatial frequencies with half-wavelength antenna spacing at the $q$th UD (the BS), respectively.
 Here, $\theta_{q,l}^{\rm UD}$ ($\theta_{q,l}^{\rm BS}$) and $\varphi_{q,l}^{\rm UD}$
 ($\varphi_{q,l}^{\rm BS}$) are the downlink horizontal and vertical AoAs (AoDs) of the
 $l$th MPC associated with the UPA, respectively. The array response vector at UD is
 given by $\bm{a}_{\rm UD}\big(\mu_{q,l}^{\rm UD},\nu_{q,l}^{\rm UD}\big)=
 \bm{a}_{\rm v}\big(\nu_{q,l}^{\rm UD}\big)\otimes \bm{a}_{\rm h}\big(\mu_{q,l}^{\rm UD}\big) \in
 \mathbb{C}^{N_{\rm UD}}$ \cite{TVT_HuChen18,2D_U_ESPRIT96,WCL_Mao18}, in which
\begin{align} 
 \bm{a}_{\rm h}\left(\mu_{q,l}^{\rm UD}\right) =& {\textstyle{1 \over {\sqrt{N_{\rm UD}^{\rm h}}}}}\big[ 1 ~
  e^{\textsf{j} \mu_{q,l}^{\rm UD}} \cdots e^{\textsf{j}\left( N_{\rm UD}^{\rm h} - 1\right)
  \mu_{q,l}^{\rm UD}}\big]^{\rm T} ,\label{eq_steering_miu} \\ 
 \bm{a}_{\rm v}\left(\nu_{q,l}^{\rm UD}\right) =& {\textstyle{1 \over {\sqrt{N_{\rm UD}^{\rm v}}}}}\big[ 1 ~
  e^{\textsf{j} \nu_{q,l}^{\rm UD}} \cdots e^{\textsf{j}\left( N_{\rm UD}^{\rm v} - 1\right)
  \nu_{q,l}^{\rm UD}}\big]^{\rm T} . \label{eq_steering_niu} 
\end{align}
 are the steering vectors associated with the horizontal and vertical directions,
 respectively. Similarly, the array response vector at BS is given by
 $\bm{a}_{\rm BS}\big(\mu_{q,l}^{\rm BS},\nu_{q,l}^{\rm BS}\big)=
 \bm{a}_{\rm v}\big(\nu_{q,l}^{\rm BS}\big) \otimes \bm{a}_{\rm h}\big(\mu_{q,l}^{\rm BS}\big)
 \in \mathbb{C}^{N_{\rm BS}}$, where the horizontal and vertical direction steering
 vectors $\bm{a}_{\rm h}\big(\mu_{q,l}^{\rm BS}\big)\in\mathbb{C}^{N_{\rm BS}^{\rm h}}$
 and $\bm{a}_{\rm h}\big(\nu_{q,l}^{\rm BS}\big)\in\mathbb{C}^{N_{\rm BS}^{\rm v}}$
 are given respectively by substituting $\mu_{q,l}^{\rm UD}$ and $N_{\rm UD}^{\rm h}$
 with $\mu_{q,l}^{\rm BS}$ and $N_{\rm BS}^{\rm h}$ in (\ref{eq_steering_miu}) as well
 as by substituting $\nu_{q,l}^{\rm UD}$ and $N_{\rm UD}^{\rm v}$ with $\nu_{q,l}^{\rm BS}$
 and $N_{\rm BS}^{\rm v}$ in (\ref{eq_steering_niu}).

 The frequency-domain channel matrix $\bm{H}_q[k]$ at the $k$th subcarrier can then be
 expressed as
\begin{equation}\label{frequency_domain_channel} 
\begin{split}
 \bm{H}_q[k]& =\beta_q \sum\limits_{l=1}^{L_q} \bm{H}_{q,l} e^{-\textsf{j}
  \frac{2\pi k f_s \tau_{q,l}}{K}}\\[-1mm]
  = \beta_q& \sum\limits_{l=1}^{L_q} \alpha_{q,l}
  \bm{a}_{\rm UD}\left(\mu_{q,l}^{\rm UD},\nu_{q,l}^{\rm UD}\right)
  \bm{a}_{\rm BS}^{\rm H}\left(\mu_{q,l}^{\rm BS},\nu_{q,l}^{\rm BS}\right) e^{-\textsf{j}
  \frac{2\pi k f_s \tau_{q,l}}{K}} ,
\end{split}
\end{equation}
 where $f_s=1/T_s$ denotes the system bandwidth, and $T_s$ is the sampling period.
 The derivation of the first equation in (\ref{frequency_domain_channel})  is shown
 in Appendix. Observe that $\bm{H}_q[k]$ does not depend on the PSF, and it exhibits
 the sparsity in delay domain due to small $L_q$ but large normalized delay spread.
 Recall that the existing CS-based solutions of \cite{JSAC_OMP17,JSAC_YangF17} have
 to estimate the effective delay-domain CIRs that include the PSF, and this PSF will
 destroy the delay-domain sparsity of mmWave channels when the order of PSF is large.
 $\bm{H}_q[k]$ in (\ref{frequency_domain_channel}) can be rewritten as
\begin{equation}\label{H_k} 
 \bm{H}_q[k] = \bm{A}_{{\rm UD},q}\bm{D}_q[k]\bm{A}_{{\rm BS},q}^{\rm H} ,
\end{equation}
 where $\bm{D}_q[k]\! =\! \text{diag}\left(\bm{d}_q[k]\right)\! \in\!
 \mathbb{C}^{L_q\times L_q}$ is the diagonal matrix in which $\bm{d}_q[k]\! =\!
 \text{diag}\left(\bm{\alpha}_q\right)\bm{\tau}_q[k]$ with $\bm{\alpha}_q\! =\!
 \beta_q\left[\alpha_{q,1}\cdots \alpha_{q,L_q}\right]^{\rm T}$ and
 $\bm{\tau}_q[k]\! =\!\big[ e^{-\textsf{j}2\pi k f_s\tau_{q,1}/K}\cdots
 e^{-\textsf{j}2\pi k f_s\tau_{q,L_q}/K}\big]^{\rm T}$, and the array response
 matrix associated with the AoAs of the $q$th UD $\bm{A}_{{\rm UD},q}\! \in\!
 \mathbb{C}^{N_{\rm UD}\times L_q}$ can be expressed as $\bm{A}_{{\rm UD},q}\! =\!
 \bm{A}_{{\rm UD},q}^{\nu}\! \odot\! \bm{A}_{{\rm UD},q}^{\mu}$ in which
 $\bm{A}_{{\rm UD},q}^{\mu}\! =\! \big[ \bm{a}_{\rm h}(\mu_{q,1}^{\rm UD})\cdots
 \bm{a}_{\rm h}(\mu_{q,L_q}^{\rm UD})\big]\! \in\! \mathbb{C}^{N_{\rm UD}^{\rm h}
 \times L_q}$ and $\bm{A}_{{\rm UD},q}^{\nu}\! =\! \big[\bm{a}_{\rm v}(\nu_{q,1}^{\rm UD})
 \cdots \bm{a}_{\rm v}(\nu_{q,L_q}^{\rm UD})\big]\! \in\!
 \mathbb{C}^{N_{\rm UD}^{\rm v}\times L_q}$ are the steering matrices corresponding to
 the horizontally and vertically spatial frequencies, respectively, while
 $\bm{A}_{{\rm BS},q}\! =\! \bm{A}_{{\rm BS},q}^{\nu}\! \odot\! \bm{A}_{{\rm BS},q}^{\mu}
 \! \in\! \mathbb{C}^{N_{\rm BS}\times L_q}$ is the array response matrix associated with
 the AoDs in which the steering matrices $\bm{A}_{{\rm BS},q}^{\mu}\! \in\!
 \mathbb{C}^{N_{\rm BS}^{\rm h}\times L_q}$ and $\bm{A}_{{\rm BS},q}^{\nu}\! \in\!
 \mathbb{C}^{N_{\rm BS}^{\rm v}\times L_q}$ have the similar form as
 $\bm{A}_{{\rm UD},q}^{\mu}$ and $\bm{A}_{{\rm UD},q}^{\nu}$, respectively.

\subsection{Obtain Horizontal/Vertical AoAs at UD}\label{S2.2}

 The downlink CE corresponds to Step\,1 to Step\,4 of Fig.~\ref{FIG2}, where the
 horizontal and vertical AoAs are estimated. We first assume that the training
 signal $\bm{s}[i,m]$ is independent of subcarriers, and its $j_1$th element can
 be designed as $\left[ \bm{s}[i,m] \right]_{j_1}\! =\! \frac{1}{N_{\rm s}^d} e^{\textsf{j}2\pi \phi_{j_1}}$
 with $\phi_{j_1}$ randomly and uniformly selected from the interval $[0, \, 1]$,
 i.e., $\phi_{j_1} \sim {\cal U}[0, ~ 1]$. Second, a predefined frequency-domain
 scrambling code $\bm{x}_d \in \mathbb{C}^{K}$ with its $k$th element being
 $x_d[k]$ for $0\! \le\! k\! \le\! K-1$ can be introduced to effectively avoid
 the high peak-to-average power ratio (PAPR) resulted from the same training signal
 used at all subcarriers\footnote{Each element in the predefined scrambling code
 $\bm{x}_d$ should satisfy $x_d^*[k]x_d[k]\! =\! 1, 0\! \le\! k\! \le\! K-1$. To
 achieve the low PAPR of training signals, we can adopt the constant-module Zadoff-Chu
 sequence as the scrambling code $\bm{x}_d$.}. Then, we can obtain the scrambled training
 signal $\bm{s}[k,i,m]$ at the $k$th subcarrier, i.e., $\bm{s}[k,i,m]\! =\! x_d[k]\bm{s}[i,m]$.
 The signals received at the UD will be first descrambled by multiplying the conjugate
 of scrambling code $\bm{x}_d^*$, which indicates that the scrambling code $\bm{x}_d$
 does not affect the subsequent signal processing. Moreover, the same digital transmit
 precoding/receive combining matrices are adopted at every subcarrier, i.e.,
 $\bm{F}_{{\rm BB},d}[k,m]\! =\! \bm{F}_{{\rm BB},d}[m]$ and $\bm{W}_{{\rm BB},q}[k,m]\! =\! \bm{W}_{{\rm BB},q}[m]$,
 for $0\! \le\! k\! \le\! K-1$. The number of independent signal streams associated
 with each subcarrier in each OFDM symbol is $N_{\rm s}^d\! \le\! N_{\rm UD}^{\rm RF}$.
 We can visualize a low-dimensional digital $M_{\rm UD}^{\rm h}\! \times\!
 M_{\rm UD}^{\rm v}$ sub-UPA, in which $M_{\rm UD}^{\rm h}$ and $M_{\rm UD}^{\rm v}$
 are the numbers of antennas in horizontal and vertical directions, from the
 high-dimensional hybrid analog/digital ${N_{\rm UD}^{\rm{h}}}\times
 {N_{\rm UD}^{\rm{v}}}$ UPA. Given $N_{\rm UD}^{\rm sub}\! =\!
 M_{\rm UD}^{\rm h} M_{\rm UD}^{\rm v}$, the BS only requires $N_d\! =\!
 \left\lceil  N_{\rm UD}^{\rm sub}/N_{\rm s}^d\right\rceil$ time slots to
 broadcast training signals, with each time slot containing $N_{\rm o}^d$ OFDM
 symbols. The choice of $M_{\rm UD}^{\rm h}$, $M_{\rm UD}^{\rm v}$ and
 $N_{\rm o}^d$ trades off estimation accuracy with training overhead\footnote{In
 this paper, the training overhead is defined as the number of OFDM symbols
 required at the CE stage. In terms of downlink CE stage, the training duration
 is $N_dN_{\rm o}^d$ OFDM symbols.}, because larger $M_{\rm UD}^{\rm h}$,
 $M_{\rm UD}^{\rm v}$ and $N_{\rm o}^d$ lead to better estimation accuracy but
 higher training overhead, and vice versa. Since the signals received by all UDs
 have the same form, we can focus on the $q$th UD and the user index $q$ can be
 omitted from $\bm{y}_q[k,i,m]$, $\bm{W}_{d,q}[m]$, $\bm{H}_q[k]$, $\bm{n}_q[k,i,m]$,
 $\bm{A}_{{\rm UD},q}$, $\bm{D}_q[k]$, $\bm{A}_{{\rm BS},q}$ and other relevant
 variables for clarity.

 By collecting the received signals of (\ref{eq_y}) associated with the $k$th subcarrier
 over all the $N_{\rm o}^d$ OFDM symbols of the $m$th time slot into the signal matrix
 $\bm{Y}_m[k]\! =\! \left[ \bm{y}[k,1,m]\cdots \bm{y}[k,N_{\rm o}^d,m]\right]
 \in \mathbb{C}^{N_{\rm s}^d\times N_{\rm o}^d}$, we have
\begin{equation}\label{Y_D_k} 
 \bm{Y}_m[k]\! =\! x_d^*[k]\bm{W}_d^{\rm H}[m] \bm{H}[k] \bm{F}_d[m] \bm{S}_d[k,m]
 \! +\! \bm{W}_d^{\rm H}[m] \bm{N}_m[k] ,
\end{equation}
 where $\bm{S}_d[k,m]\! =\! \left[\bm{s}[k,1,m]\cdots\bm{s}[k,N_{\rm o}^d,m]\right]\!
 =\! x_d[k]\bm{S}_d[m]$ with $\bm{S}_d[m]\! =\! \left[\bm{s}[1,m]\cdots\bm{s}
 [N_{\rm o}^d,m]\right]\! \in\! \mathbb{C}^{N_{\rm s}^d\times N_{\rm o}^d}$,
 and $\bm{N}_m[k]\! =\! \left[\bm{n}[k,1,m] \cdots\bm{n}[k,N_{\rm o}^d,m]\right]\!
 \in\! \mathbb{C}^{N_{\rm UD}\times N_{\rm o}^d}$. Since the BS transmits the
 common random signal $\bm{F}_d[m] \bm{S}_d[m]$, the transmit precoding matrix
 $\bm{F}_d[m]\! =\! \bm{F}_{{\rm RF},d}[m] \bm{F}_{{\rm BB},d}[m]$ should be a
 random matrix. This is achieved by designing $\bm{F}_{{\rm RF},d}[m]$ as
 $\left[\bm{F}_{{\rm RF},d}[m]\right]_{j_1,j_2}\!\! =\! \frac{1}{\sqrt{N_{\rm BS}}}
 e^{\textsf{j} \vartheta_{3,j_1,j_2}}$ with ${\vartheta_{3,j_1,j_2}}$ randomly
 and uniformly selected from ${\cal A}$, and designing $\bm{F}_{{\rm BB},d}[m]$ as
 $\left[ \bm{F}_{{\rm BB},d}[m]\right]_{j_1,j_2} \!\! =\! e^{\textsf{j}2\pi a_{j_1,j_2}}$
 with $a_{j_1,j_2}\! \sim\! {\cal U}\left[0,1\right]$. The BS can use the same
 transmit precoding matrix $\bm{F}_d\!=\!\bm{F}_d[m]$ to send the same sounding
 signal $\bm{S}_d\! =\! \bm{S}_d[m]$ for every time slot. By stacking the received
 signal matrices $\bm{Y}_m[k]$ of (\ref{Y_D_k}) over the $N_d$ time slots into
 $\widetilde{\bm{Y}}_d[k]\! =\! \left[ \bm{Y}_1^{\rm T}[k]\cdots \bm{Y}_{N_d}^{\rm T}[k]
 \right]^{\rm T}\! \in\! \mathbb{C}^{N_d N_{\rm s}^d \times N_{\rm o}^d}$, we have
\begin{equation}\label{Y_D_k_tilde} 
 \widetilde{\bm{Y}}_d[k]\! =\! \widetilde{\bm{W}}_d^{\rm H} \bm{A}_{\rm UD} \bm{D}[k]
 \bm{A}_{\rm BS}^{\rm H} \bm{F}_d \bm{S}_d\! +\! \text{Bdiag}\left(\check{\bm{W}}_d\right)
 \widetilde{\bm{N}}_d[k] ,
\end{equation}
 where $\widetilde{\bm{W}}_d\! =\! \left[\bm{W}_d[1]\cdots \bm{W}_d[N_d]\right]\! \in\!
 \mathbb{C}^{N_{\rm UD}\times N_d N_{\rm s}^d}$ aggregates the downlink receive
 combining matrices used in the $N_d$ time slots, and $\text{Bdiag}\left(
 \check{\bm{W}}_d\right)\! =\!\text{Bdiag}\left(\left[\bm{W}_d^{\rm H}[1]\cdots
 \bm{W}_d^{\rm H}[N_d]\right]\right)\! \in\! \mathbb{C}^{N_d N_{\rm s}^d \times
 N_d N_{\rm UD}}$, while $\widetilde{\bm{N}}_d[k] \! =\! \left[\bm{N}_1^{\rm T}[k]
 \cdots \bm{N}_{N_d}^{\rm T}[k]\right]^{\rm T}\! \in \!
 \mathbb{C}^{N_d N_{\rm UD}\times N_{\rm o}^d}$ is the corresponding noise matrix.

 Multiplying $\widetilde{\bm{Y}}_d[k]$ with $\bm{J}_d\! =\! \big[
 \bm{I}_{N_{\rm UD}^{\rm sub}} \, \bm{O}_{N_{\rm UD}^{\rm sub}\times
 \left( N_d N_{\rm s}^d - N_{\rm UD}^{\rm sub}\right)} \big]$ and aggregating the
 resulting signals over all the $K$ subcarriers lead to the signal matrix
 $\bar{\bm{Y}}_d\! =\! \big[ \bm{J}_d \widetilde{\bm{Y}}_d[0] ~ \bm{J}_d \widetilde{\bm{Y}}_d[1]
  \cdots \bm{J}_d \widetilde{\bm{Y}}_d[K-1] \big]\! \in\!
 \mathbb{C}^{N_{\rm UD}^{\rm sub} \times K N_{\rm o}^d}$ as
\begin{equation}\label{Y_bar_D} 
 \bar{\bm{Y}}_d = \bar{\bm{A}}_{\rm UD} \bar{\bm{S}}_d
  + \bar{\bm{N}}_d,
\end{equation}
 where $\bar{\bm{N}}_d\! =\! \bm{J}_d\text{Bdiag}\left(\check{\bm{W}}_d\right)
 \big[\widetilde{\bm{N}}_d[0] ~ \widetilde{\bm{N}}_d[1]\cdots \widetilde{\bm{N}}_d[K-1]\big]$,
 $\bar{\bm{A}}_{\rm UD}\! =\! \bm{J}_d\widetilde{\bm{W}}_d^{\rm H}\bm{A}_{\rm UD}$, and
 $\bar{\bm{S}}_d\! =\! \left[ \bar{\bm{S}}_d[0] ~ \bar{\bm{S}}_d[1]\cdots
 \bar{\bm{S}}_d[K-1]\right]$ with $\bar{\bm{S}}_d[k]\! =\! \bm{D}[k]\bm{A}_{\rm BS}^{\rm H}
 \bm{F}_d\bm{S}_d$.
 Observe from (\ref{Y_D_k_tilde}) and (\ref{Y_bar_D}) that we cannot directly apply
 powerful spectrum estimation techniques \cite{2D_U_ESPRIT96,SSD_R_U_ESPRIT98} to
 estimate the horizontal/vertical AoAs from $\bar{\bm{Y}}_d$, since the shift-invariance
 structure of the array response matrix $\bm{A}_{\rm UD}$ does not hold in hybrid receive
 array \cite{TSP_JADE98,TSP_Vandermonde98}. We propose to visualize the high-dimensional
 hybrid array as a low-dimensional digital array by designing appropriate aggregated receive
 combining matrix $\widetilde{\bm{W}}_d$ so that the shift-invariance structure of array
 response can be reconstructed and therefore super-resolution CE based on spectrum estimation
 techniques can be harnessed.

\subsection{Design Receive Combining Matrix at UD}\label{S2.3}

 Without loss of generality, we consider $N_{\rm s}^d\! =\! N_{\rm UD}^{\rm RF}\! -\! 1$
 independent signal streams. First, we utilize a unitary matrix $\bm{U}_{N_{\rm UD}^{\rm RF}}
 \! =\! \big[ \bm{u}_1 \cdots \bm{u}_{N_{\rm UD}^{\rm RF}} \big]\! \in\!
 \mathbb{C}^{N_{\rm UD}^{\rm RF} \times N_{\rm UD}^{\rm RF}}$ to design the digital
 receive combining matrix $\bm{W}_{\rm BB}[m]\! \in\!
 \mathbb{C}^{N_{\rm UD}^{\rm RF}\times N_{\rm s}^d}$ of the $m$th time slot's receive
 combining matrix $\bm{W}_d[m]\! =\! \bm{W}_{\rm RF}[m] \bm{W}_{\rm BB}[m]$ for
 $1\! \le\! m\! \le\! N_d$. Specifically, $\bm{W}_{\rm BB}[m]\! =\!
 \bm{U}_{N_{\rm UD}^{\rm RF}[:,1:N_{\rm s}^d]}$. To design the analog receive combining
 matrix $\bm{W}_{\rm RF}[m]\! \in\! \mathbb{C}^{N_{\rm UD}\times N_{\rm UD}^{\rm RF}}$,
 we construct the matrix $\bm{\varXi}_d\! \in\! \mathbb{R}^{N_{\rm UD}\times N_d N_{\rm s}^d}$ as
\begin{equation}\label{Xi_D_Index_Matrix} 
\begin{split}
 \bm{\varXi}_d &= \left[\!\! \begin{array}{c}
  \bm{I}_{M_{\rm UD}^{\rm v} + 1} \otimes \bm{B} \\
  \bm{O}_{\left( N_{\rm UD} - N_{\rm UD}^{\rm h}\left( M_{\rm UD}^{\rm v} + 1\right)\right)
  \times M_{\rm UD}^{\rm h}\left( M_{\rm UD}^{\rm v} + 1\right)} \end{array} \right]\\
  &\quad \times \left[\!\! \begin{array}{c} \bm{I}_{N_{\rm s}^d N_d} \\
  \bm{O}_{\left( M_{\rm UD}^{\rm h}\left( M_{\rm UD}^{\rm v} + 1\right) - N_{\rm s}^d N_d\right)
  \times N_{\rm s}^d N_d} \end{array} \right] ,
\end{split}
\end{equation}
 where $\bm{B}\! =\! \big[ \bm{I}_{M_{\rm UD}^{\rm h}} \, \bm{O}_{M_{\rm UD}^{\rm h}
 \times (N_{\rm UD}^{\rm h} - M_{\rm UD}^{\rm h})}\big]^{\rm T} \! \in\!
 \mathbb{R}^{N_{\rm UD}^{\rm h}\times M_{\rm UD}^{\rm h}}$. Then we take the sub-matrix
 $\bm{\varXi}_{d,m}^{\rm sub}\! =\! \bm{\varXi}_{d\,\left[:,(m-1)N_{\rm s}^d+1:m N_{\rm s}^d\right]}
 \! \in\! \mathbb{R}^{N_{\rm UD}\times N_{\rm s}^d}$ and define $\bar{\bm{\xi}}_{d,m}\! =\!
 \text{vec}({\bm{\varXi}_{d,m}^{\rm sub}})$ to construct the ordered index set ${\cal D}_m\!
 =\! {\rm find}\left( \bar{\bm{\xi}}_{d,m}\! \ne\! 0\right)$ with $|{\cal D}_m|_c = N_{\rm s}^d$.
 Next we perform the modulo operation on ${\cal D}_m$ with $N_{\rm UD}$ to get the
 ordered index set ${\cal I}_m\! =\! \text{mod}( {\cal D}_m,N_{\rm UD})$ with
 $|{\cal I}_m|_c = N_{\rm s}^d$. The rows of $\bm{W}_{\rm RF}[m]$ whose indices
 correspond to ${\cal I}_m$ are determined by $\bm{W}_{\rm BB}[m]$ as
 $\bm{W}_{\rm RF}[m]_{[{\cal I}_m,:]}\! =\! \bm{W}_{\rm BB}^{\rm H}[m]$, while the rest
 rows of $\bm{W}_{\rm RF}[m]$ consist of the $(N_{\rm UD}\! - \! N_{\rm s}^d)$ identical
 $\bm{u}_{N_{\rm UD}^{\rm RF}}^{\rm H}$. The phase value of arbitrary element in the
 designed $\bm{W}_{\rm RF}[m]$, denoted by $\vartheta_d$, is then quantized to $\vartheta
 \! \in \! {\cal A}$ by minimizing the Euclidean distance according to
 $\arg \min\nolimits_{\vartheta \in {\cal A}} \! \left\| \vartheta_d\! -\!
 \vartheta\right\|_2$. Thus, the $m$th receive combining matrix can be obtained as
 $\bm{W}_d[m]\! =\! \bm{W}_{\rm RF}[m]\bm{W}_{\rm BB}[m]$ for $1\le m\le N_d$.

\begin{algorithm}[tp!]
\caption{Proposed Receive Combining Matrix Design}
\label{ALG1}
\begin{algorithmic}[1]
\REQUIRE $N_d$, $N_{\rm s}^d$, $N_{\rm UD}^{\rm RF}$, $N_{\rm UD}^{\rm h}$, $N_{\rm UD}^{\rm v}$,
  $M_{\rm UD}^{\rm h}$, $M_{\rm UD}^{\rm v}$
\ENSURE $\widetilde{\bm{W}}_d$
\STATE Generate unitary matrix $\bm{U}_{N_{\rm UD}^{\rm RF}} = \big[ \bm{u}_1 \cdots
  \bm{u}_{N_{\rm UD}^{\rm RF}} \big]$
\STATE Construct index matrix $\bm{\varXi}_d$ of (\ref{Xi_D_Index_Matrix})
\FOR{$m = 1, 2, \cdots , N_d$}
  \STATE $\bm{W}_{\rm BB}[m] = \bm{U}_{N_{\rm UD}^{\rm RF}[:,1:N_{\rm s}^d]}$, and
    initialize $\bm{W}_{\rm RF}[m] = \bm{1}_{N_{\rm UD}} \otimes \bm{u}_{N_{\rm UD}^{\rm RF}}^{\rm H}$
  \STATE Extract $\bm{\varXi}_{d,m}^{\rm sub} = \bm{\varXi}_{d\left[:,(m-1)N_{\rm s}^d+1:m N_{\rm s}^d\right]}$,
    and obtain $\bar{\bm{\xi}}_{d,m} = \text{vec}(\bm{\varXi}_{d,m}^{\rm sub})$
  \STATE Obtain ordered index set ${\cal I}_m = \text{mod}\left( \text{find}\left(
   \bar{\bm{\xi}}_{d,m} \ne 0 \right),N_{\rm UD} \right)$
\STATE Replace $\bm{W}_{\rm RF}[m]_{[{\cal I}_m,:]} \leftarrow  \bm{W}_{\rm BB}^{\rm H}[m]$
\STATE Quantize phase values of $\bm{W}_{\rm RF}[m]$ based on ${\cal A}$
\STATE $\bm{W}_d[m] = \bm{W}_{\rm RF}[m]\bm{W}_{\rm BB}[m]$
\ENDFOR
\RETURN $\widetilde{\bm{W}}_d = \left[ \bm{W}_d[1] \cdots \bm{W}_d[N_d] \right]$
\end{algorithmic}
\end{algorithm}

 The proposed design for $\widetilde{\bm{W}}_d$ is summarized in Algorithm~\ref{ALG1}.
 Since the number of RF chains is usually the power of 2, we can adopt Hadamard matrix
 for $N_q^{\rm ps}\! \ge\! 1$ or discrete Fourier transform (DFT) matrix for $N_q^{\rm ps}\!
 \ge\! 2$ to construct $\bm{U}_{N_{\rm UD}^{\rm RF}}$\footnote{Note that the phase value of
 every entry of the quantized Hadamard or DFT matrices still belongs to the set ${\cal A}$,
 and therefore we ensure the columns of the selected ${{\bm{U}}_{N_{{\rm{UD}}}^{{\rm{RF}}}}}$
 to be mutually orthogonal.}. Clearly, our design can be used for the PSN with arbitrary
 $N_q^{\rm ps}$, even the extremely low resolution PSN with $N_q^{\rm ps}\! =\! 1$. With
 the designed $\widetilde{\bm{W}}_d$, we have
\begin{equation}\label{Y_bar_D_2} 
 \bar{\bm{A}}_{\rm UD} = \bm{J}_d \widetilde{\bm{W}}_d^{\rm H}\left(\bm{A}_{\rm UD}^{\nu}
  \odot \bm{A}_{\rm UD}^{\mu}\right) = \bar{\bm{A}}_{\rm UD}^{\nu} \odot \bar{\bm{A}}_{\rm UD}^{\mu} .
\end{equation}
 Clearly, $\bar{\bm{A}}_{\rm UD}^{\mu}\! \in\! \mathbb{C}^{M_{\rm UD}^{\rm h} \times L}$
 ($\bar{\bm{A}}_{\rm UD}^{\nu}\! \in\! \mathbb{C}^{M_{\rm UD}^{\rm v} \times L}$) is the
 matrix containing the first $M_{\rm UD}^{\rm h}$ ($M_{\rm UD}^{\rm v}$) rows of
 $\bm{A}_{\rm UD}^{\mu}$ ($\bm{A}_{\rm UD}^{\nu}$). Thus, $\bar{\bm{A}}_{\rm UD}$
 maintains the double shift-invariance structure of the original array response matrix
 $\bm{A}_{\rm UD}$ for both horizontal and vertical AoAs \cite{TSP_Vandermonde98},
 and the designed $\widetilde{\bm{W}}_d$ can be used to visualize the high-dimensional
 hybrid analog/digital array as a low-dimensional digital array. Therefore, we can
 utilize the MDU-ESPRIT algorithm detailed in Section~\ref{S4} to obtain the super-resolution
 estimates of horizontal/vertical AoAs at UD. Since the ESPRIT-type algorithms \cite{2D_U_ESPRIT96,TSP_JADE98,TSP_Vandermonde98,SSD_R_U_ESPRIT98} require the
 knowledge of the number of MPCs, we next turn to the task of acquiring the number
 of MPCs at the receiver, i.e., Step~2 of Fig.~\ref{FIG2}.

\subsection{EVD-Based Estimate for Number of MPCs}\label{S2.4}

 In OFDM systems, the channels of multiple adjacent subcarriers within coherence
 bandwidth are highly correlated. If the maximum delay spread is
 $\tau_{\max}\! =\! N_{\rm c} T_{\rm s}$ with $N_{\rm c}$ delay taps, the channel coherence
 bandwidth is $B_c\! \approx\! \frac{1}{\tau_{\max}}\! =\! \frac{f_{\rm s}}{N_{\rm c}}$. Then
 we can jointly use the measurements of $P\! \le\! \frac{B_c}{\Delta f}\! =\! \frac{K}{N_{\rm c}}$
 adjacent subcarriers to estimate the number of MPCs, where $\Delta f\! =\! \frac{f_{\rm s}}{K}$
 is the subcarrier's bandwidth. Specifically, by dividing $K$ signal matrices
 $\big\{ \bm{J}_d \widetilde{\bm{Y}}_d[k] \big\}_{k=0}^{K-1}$ into $N_P\! =\!
 \left\lfloor K/P \right\rfloor$ groups, we can obtain the $n_p$th measurement matrix
 $\check{\bm{Y}}_d[n_p]\! \in\! \mathbb{C}^{N_{\rm UD}^{\rm sub} \times N_{\rm o}^d}$,
 as the average of the measurements in the $n_p$th group
\begin{equation}\label{Y_d_tilde_np} 
 \check{\bm{Y}}_d[n_p] = \frac{1}{P}\sum\nolimits_{k=(n_p - 1)P}^{n_p P - 1}
  \bm{J}_d \widetilde{\bm{Y}}_d[k] , ~ 1\le n_p\le N_P.
\end{equation}
 The $N_P$ average measurements are collected as $\check{\bm{Y}}_d\! =\!
 \left[ \check{\bm{Y}}_d[1] \cdots \check{\bm{Y}}_d[N_P] \right]\! \in\!
 \mathbb{C}^{N_{\rm UD}^{\rm sub} \times N_{\rm o}^d N_P}$, and the covariance matrix
 of $\check{\bm{Y}}_d$ is $\bm{R}_d\! =\! \frac{1}{N_{\rm o}^d N_P} \check{\bm{Y}}_d
 \check{\bm{Y}}_d^{\rm H}$. According to the eigenvalue decomposition (EVD), we
 obtain $\bm{R}_d\! =\! \left[ \bm{U}_s ~ \bm{U}_n \right]\text{diag}\left(
 \bm{\lambda}_d \right)\left[ \bm{U}_s ~ \bm{U}_n \right]^{\rm H}$, where $\bm{\lambda}_d
 \! =\!\big[ \lambda_1 \cdots \lambda_L ~ \lambda_{L+1} \cdots \lambda_{N_{\rm UD}^{\rm sub}}
 \big]^{\rm T}$ $\! =\! \left[ \bm{\lambda}_s^{\rm T} ~ \bm{\lambda}_n^{\rm T}\right]^{\rm T}$
 is the eigenvalue vector with the eigenvalues arranged in descending order, $\bm{U}_s$ and
 $\bm{U}_n$ are the eigenvector matrices corresponding to the signal and noise subspaces,
 respectively, while $\bm{\lambda}_s\! =\! \left[ \lambda_1 \cdots \lambda_L \right]^{\rm T}$
 and $\bm{\lambda}_n\! =\! \big[ \lambda_{L+1}\cdots \lambda_{N_{\rm UD}^{\rm sub}}\big]^{\rm T}$
 are the eigenvalue vectors related to $\bm{U}_s$ and $\bm{U}_n$, respectively.
 The number of MPCs $L$ is the dimension of $\bm{\lambda}_s$.

 To obtain an accurate estimate of $L$, we first construct $\widetilde{\bm{\lambda}}\! =\!
 [\bm{\lambda}_s^{\rm T} ~ \bm{0}_{N_{\rm UD}^{\rm sub}-L}^{\rm T}]^{\rm T}\! \in\!
 \mathbb{C}^{N_{\rm UD}^{\rm sub}}$. The optimal estimate of $\widetilde{\bm{\lambda}}$
 can be acquired by solving the following optimization problem
\begin{equation}\label{Optimization_Soft} 
 \widetilde{\bm{\lambda}}^{\star} = \arg \min\limits_{ \widetilde{\bm{\lambda}}\ge
  \bm{0}_{N_{\rm UD}^{\rm sub}}} \frac{1}{2}
  \big\| \widetilde{\bm{\lambda}} - \bm{\lambda}_d \big\|_2^2 + \varepsilon \big\|
  \widetilde{\bm{\lambda}} \big\|_1 ,
\end{equation}
 where $\varepsilon$ is the threshold parameter related to the AWGN power, which is
 determined experimentally. Clearly, the solution to the optimization problem
 (\ref{Optimization_Soft}) is \cite{Soft95}
\begin{equation}\label{Soft_Function} 
 \widetilde{\lambda}_i^{\star} =\left\{ \begin{array}{cl}
  \lambda_i - \varepsilon , & \lambda_i \ge \varepsilon  \\
  0 , & \lambda_i < \varepsilon ,
\end{array} \right.
\end{equation}
 where $\widetilde{\lambda}_i^{\star}$ is the $i$th element of $\widetilde{\bm{\lambda}}^{\star}$.
 From the estimate $\widetilde{\bm{\lambda}}^{\star}$, we obtain the estimate of the number
 of MPCs, denoted by $\widehat{L}$, which is the input to the MDU-ESPRIT algorithm for
 estimating $\widehat{L}$ pairs of horizontal and vertical AoAs. The resulting estimates
 $\big\{\widehat{\theta}_l^{\rm UD},\widehat{\varphi}_l^{\rm UD}\big\}_{l=1}^{\widehat{L}}$
 are quantized as $\left\{ \bar{\theta}_l^{\rm UD},\bar{\varphi}_l^{\rm UD}\right\}_{l=1}^{\widehat{L}}$
 with $N_q^{\rm ang}$ angle quantized bits in $[-\pi/2, ~ \pi/2]$.

 Finally, only the few bits of the quantized angle estimates are fed back to BS through
 the low-frequency control link with limited resource [2]. Thus, since very little data
 needs to be transmitted via the feedback link, the feedback overhead at the AoAs feedback
 stage can be ignored in our proposed closed-loop sparse CE scheme\footnote{In the open-loop
 CE schemes \cite{JSAC_OMP17,TWC_SW_OMP18}, the support sets and channel gains for every
 subcarrier estimated at the receiver also need to be fed back to transmitter to perform
 the following signal processing such as beamforming design or channel equalization
 \cite{Gao_UDN15,JSTSP_Heath16}. Compared with these schemes, our proposed closed-loop
 CE scheme only feeds back the dominant channel parameters estimated by the BS and UD
 to each other, and thus, its feedback overhead is almost negligible.}.

\section{Uplink Channel Estimation Stage}\label{S3}

\subsection{Obtain Horizontal/Vertical AoDs and Delays at BS}\label{S3.1}

 At the uplink CE stage, the BS jointly estimates the horizontal/vertical AoDs
 and delays for each UD. Due to the channel reciprocity in TDD systems
 \cite{JSAC_Yuan17,WC_GaoZ18,Tcom_ZhangS19}, the uplink channel matrix for the
 $q$th UD is given by $\bm{H}^{\rm T}[k]\! =\! \bm{A}_{\rm BS}^* \bm{D}[k]
 \bm{A}_{\rm UD}^{\rm T}\! \in\! \mathbb{C}^{N_{\rm BS} \times N_{\rm UD}}$, where
 again the user index $q$ is omitted. We employ $N_{\rm s}^u\! =\! N_{\rm BS}^{\rm RF}
 \! -\! 1$ independent signal streams, and a low-dimensional digital $M_{\rm BS}^{\rm h}
 \times M_{\rm BS}^{\rm v}$ sub-UPA with $M_{\rm BS}^{\rm h}$ and $M_{\rm BS}^{\rm v}$
 antennas in horizontal and vertical directions is visualized from the high-dimensional
 hybrid analog/digital ${N_{\rm BS}^{\rm h}} \times {N_{\rm BS}^{\rm v}}$ UPA at the BS.
 Each UD requires $N_u\! =\! \left\lceil N_{\rm BS}^{\rm sub}/N_{\rm s}^u \right\rceil$
 time slots with $N_{\rm BS}^{\rm sub}\! =\! M_{\rm BS}^{\rm h} M_{\rm BS}^{\rm v}$ to
 transmit the training signals, and each time slot consists of $N_{\rm o}^u$ OFDM symbols.
 Hence, the uplink CE for $Q$ UDs has a training overhead of $Q N_u N_{\rm o}^u$, and the
 total training overhead of the proposed closed-loop sparse CE scheme is $T_{\rm CE}\! =\!
 N_d N_{\rm o}^d\! +\! Q N_u N_{\rm o}^u$. Similar to (\ref{Y_D_k_tilde}),
 after the frequency-domain scrambling/descrambling operation, the signal matrix
 $\widetilde{\bm{Y}}_u[k]\! \in\! \mathbb{C}^{N_u N_{\rm s}^u \times N_{\rm o}^u}$
 received by the BS at the $k$th subcarrier and over the $N_u$ time slots can be expressed as
\begin{equation}\label{Y_U_k_tilde} 
 \widetilde{\bm{Y}}_u[k] = \widetilde{\bm{W}}_u^{\rm H} \bm{A}_{\rm BS}^* \bm{D}[k]
  \bm{A}_{\rm UD}^{\rm T} \bm{F}_u \bm{S}_u + \text{Bdiag}\left( \bar{\bm{W}}_u\right)
  \widetilde{\bm{N}}_u[k] , 
\end{equation}
 where $\widetilde{\bm{W}}_u\! =\! \left[ \bm{W}_u[1] \cdots \bm{W}_u[N_u] \right]\! \in\!
 \mathbb{C}^{N_{\rm BS} \times N_u N_{\rm s}^u}$ with $\bm{W}_u[m]\! \in\!
 \mathbb{C}^{N_{\rm BS} \times N_{\rm s}^u}$ being the uplink receive combining matrix
 used in the $m$th time slot for $1\le m\le N_u$, $\bar{\bm{W}}_u\! =\! \left[
 \bm{W}_u^{\rm H}[1] \cdots \bm{W}_u^{\rm H}[N_u] \right]$, and $\bm{F}_u\! \in\!
 \mathbb{C}^{N_{\rm UD} \times N_{\rm s}^u}$ is the multi-beam transmit precoding
 matrix at UD, while $\bm{S}_u\! \in\! \mathbb{C}^{N_{\rm s}^u \times N_{\rm o}^u}$
 is the uplink training signal matrix,
 and $\widetilde{\bm{N}}_u[k]$ is the uplink noise matrix. $\widetilde{\bm{Y}}_u[k]$ is multiplied by
 $\bm{J}_u\! =\! \big[\bm{I}_{N_{\rm BS}^{\rm sub}} ~  \bm{O}_{N_{\rm BS}^{\rm sub} \times
 ( N_{\rm s}^u N_u - N_{\rm BS}^{\rm sub}) } \big]\! \in\!
 \mathbb{R}^{N_{\rm BS}^{\rm sub} \times N_{\rm s}^u N_u}$ and the result is converted
 into the vector $\widetilde{\bm{y}}_u[k]\! =\! \text{vec}\big(\big(\bm{J}_u
 \widetilde{\bm{Y}}_u[k]\big)^{\rm T} \big)$, i.e.,
\begin{equation}\label{h_U_k_smile} 
 \widetilde{\bm{y}}_u[k] = \big( \bar{\bm{A}}_{\rm BS} \odot \left( \bm{A}_{\rm UD}^{\rm T}
  \bm{F}_u \bm{S}_u \right)^{\rm T} \big) \text{diag}(\bm{\alpha}) \bm{\tau}[k] +
  \widetilde{\bm{n}}_u[k] ,
\end{equation}
 where we have used the identity $\text{vec}(\bm{ABC})\! =\! \big( \bm{C}^{\rm T} \odot
 \bm{A} \big) \bm{b}$ with $\bm{B}\! =\! \text{diag}(\bm{b})$ \cite{Matrix13},
 $\bar{\bm{A}}_{\rm BS}\! =\! \bm{J}_u \widetilde{\bm{W}}_u^{\rm H} \bm{A}_{\rm BS}^*$,
 and $\widetilde{\bm{n}}_u[k]$ is the corresponding noise vector. Furthermore, by
 collecting $\widetilde{\bm{y}}_u[k]\! \in\! \mathbb{C}^{N_{\rm BS}^{\rm sub} N_{\rm o}^u}$
 for $0\le k\le K-1$, we obtain the aggregated signal matrix $\widetilde{\bm{Y}}_u\! =\!
 \big[ \widetilde{\bm{y}}_u[0] ~ \widetilde{\bm{y}}_u[1] \cdots \widetilde{\bm{y}}_u[K-1]
 \big] \! \in\! \mathbb{C}^{N_{\rm BS}^{\rm sub} N_{\rm o}^u \times K}$ given by
\begin{equation}\label{H_U_smile} 
 \widetilde{\bm{Y}}_u = \big( \bar{\bm{A}}_{\rm BS} \odot \left(
  \bm{A}_{\rm UD}^{\rm T} \bm{F}_u \bm{S}_u \right)^{\rm T} \big)
  \text{diag}\left( \bm{\alpha} \right) \bm{A}_{\bm{\tau}}^{\rm T} + \widetilde{\bm{N}}_u ,
\end{equation}
 where $\bm{A}_{\bm{\tau}}\! =\! \left[ \bm{\tau}[0] ~ \bm{\tau}[1]\cdots
 \bm{\tau}[K\! -\! 1]\right]^{\rm T} \! \in\! \mathbb{C}^{K \times L}$, and
 $\widetilde{\bm{N}}_u$ is the aggregated noise matrix. Recalling $\bm{\tau}[k]\! =\!
 \left[ e^{-\textsf{j}2\pi k f_s\tau_1/K} \cdots e^{-\textsf{j}2\pi k f_s\tau_L/K}
 \right]^{\rm T}$, we have $\bm{A}_{\bm{\tau}}\! =\! \left[ \bm{a}_{\tau}\left(
 \mu_1^{\tau}\right)\cdots \bm{a}_{\tau}\left( \mu_L^{\tau} \right)\right]$, in which
 $\bm{a}_{\tau}\left( \mu _l^{\tau} \right)\! =\! \left[ 1 ~ e^{\textsf{j}\mu_l^{\tau}}
 \cdots e^{\textsf{j}(K-1)\mu_l^{\tau}} \right]^{\rm T}\! \in \! \mathbb{C}^K$ with
 $\mu_l^{\tau}\! =\! -2\pi f_s\tau_l/K$. Observe that $\bm{A}_{\bm{\tau}}$ can be
 considered as the steering matrix associated with the delays $\{\tau_l\}_{l=1}^L$.
 Taking the vectorization of $\widetilde{\bm{Y}}_u$, i.e., $\check{\bm{y}}_u\! =\!
 \text{vec}\big( \widetilde{\bm{Y}}_u \big)$, leads to
\begin{equation}\label{y_U_tilde} 
 \check{\bm{y}}_u = \big( \left( \bm{A}_{\bm{\tau}} \odot \bar{\bm{A}}_{\rm BS}
 \right) \odot \left( \bm{A}_{\rm UD}^{\rm T} \bm{F}_u \bm{S}_u \right)^{\rm T}
 \big) \bm{\alpha} + \check{\bm{n}}_u ,
\end{equation}
 where we have used the identity $\bm{A}\! \odot\! \left( \bm{B}\! \odot\! \bm{C} \right)\!
 =\! \left( \bm{A}\! \odot\! \bm{B} \right)\! \odot\! \bm{C}$ \cite{{Matrix13}}, and
 $\check{\bm{n}}_u\! =\! \text{vec}\big( \widetilde{\bm{N}}_u \big)$. We further reshape
 $\check{\bm{y}}_u\! \in\! \mathbb{C}^{K N_{\rm BS}^{\rm sub} N_{\rm o}^u}$ as the matrix
 $\check{\bm{Y}}_u\! =\! \text{mat}\big(\check{\bm{y}}_u; N_{\rm o}^u,K N_{\rm BS}^{\rm sub}
 \big)\! \in\! \mathbb{C}^{N_{\rm o}^u \times K N_{\rm BS}^{\rm sub}}$:
\begin{equation}\label{Y_U_smile} 
 \check{\bm{Y}}_u = \left( \bm{A}_{\rm UD}^{\rm T} \bm{F}_u \bm{S}_u \right)^{\rm T} \text{diag}(\bm{\alpha})
  \left( \bm{A}_{\bm{\tau}} \odot \bar{\bm{A}}_{\rm BS} \right)^{\rm T} + \check{\bm{N}}_u ,
\end{equation}
 where $\check{\bm{N}}_u\!=\!\text{mat}\left(\check{\bm{n}}_u; N_{\rm o}^u,K N_{\rm BS}^{\rm sub} \right)$.
 Hence, $\bar{\bm{Y}}_u\!=\! \check{\bm{Y}}_u^{\rm T}\! \in\! \mathbb{C}^{K N_{\rm BS}^{\rm sub}
 \times N_{\rm o}^u}$ can be written as
\begin{equation}\label{Y_bar_U_1} 
 \bar{\bm{Y}}_u = \left( \bm{A}_{\bm{\tau}} \odot \bar{\bm{A}}_{\rm BS} \right)
  \text{diag}(\bm{\alpha}) \left( \bm{A}_{\rm UD}^{\rm T} \bm{F}_u \bm{S}_u \right) + \check{\bm{N}}_u^{\rm T} .
\end{equation}

\begin{figure*}[!tp]
\begin{center}
 \includegraphics[width=0.9\linewidth, keepaspectratio]{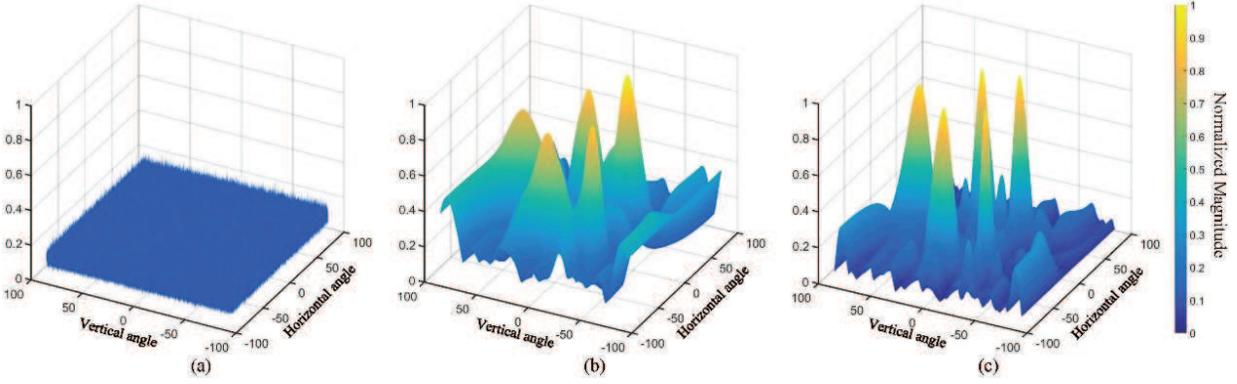}
\end{center}
\captionsetup{font = {footnotesize}, name = {Fig.}, labelsep = period}
\caption{Comparison of beam patterns, where $N_{\rm BS}^{\rm RF}=N_{\rm UD}^{\rm RF}=4$
 and the AoAs of the 5 MPCs are known to the UD: (a)~Random transmit precoding matrix
 with $8 \times 8$ UPA, (b)~Multi-beam transmit precoding matrix with $8 \times 8$ UPA,
 and (c)~Multi-beam transmit precoding matrix with $16 \times 16$ UPA.}
\label{FIG4}
\end{figure*}

 From $\bar{\bm{A}}_{\rm BS}\! =\! \bm{J}_u\widetilde{\bm{W}}_u^{\rm H}\bm{A}_{\rm BS}^*$,
 we observe that $\widetilde{\bm{W}}_u$ may destroy the shift-invariance structure of
 $\bm{A}_{\rm BS}$. Similar to downlink CE, we can design $\widetilde{\bm{W}}_u$ using
 Algorithm~\ref{ALG1} by replacing the input parameters $N_d$, $N_{\rm s}^d$,
 $N_{\rm UD}^{\rm RF}$, $N_{\rm UD}^{\rm h}$, $N_{\rm UD}^{\rm v}$, $M_{\rm UD}^{\rm h}$
 and $M_{\rm UD}^{\rm v}$ for UD with $N_u$, $N_{\rm s}^u$, $N_{\rm BS}^{\rm RF}$,
 $N_{\rm BS}^{\rm h}$, $N_{\rm BS}^{\rm v}$, $M_{\rm BS}^{\rm h}$ and $M_{\rm BS}^{\rm v}$
 for BS. By substituting the designed $\widetilde{\bm{W}}_u$ into (\ref{Y_bar_U_1}), we
 obtain
\begin{equation}\label{Y_bar_U} 
 \bar{\bm{Y}}_u = \bm{A}_{\bm{\tau}{\rm BS}} \bar{\bm{S}}_u + \check{\bm{N}}_u^{\rm T} ,
\end{equation}
 where $\bar{\bm{S}}_u\! =\! \text{diag}(\bm{\alpha})\left( \bm{A}_{\rm UD}^{\rm T} \bm{F}_u
 \bm{S}_u \right)$, and $\bm{A}_{\bm{\tau}{\rm BS}}\! \in\! \mathbb{C}^{K M_{\rm BS}^{\rm v}
 M_{\rm BS}^{\rm h} \times L}$ is given by
\begin{equation}\label{A_tau_BS_bar} 
\begin{split}
 \bm{A}_{\bm{\tau}{\rm BS}} &= \bm{A}_{\bm{\tau}} \odot \big( \bm{J}_u \widetilde{\bm{W}}_u^{\rm H}
  \left( \left(\bm{A}_{\rm BS}^{\nu}\right)^* \odot \left( \bm{A}_{\rm BS}^{\mu}\right)^*
  \right) \big)\\
  &= \bm{A}_{\bm{\tau}} \odot \bar{\bm{A}}_{\rm BS}^{\nu} \odot
  \bar{\bm{A}}_{\rm BS}^{\mu} .
\end{split}
\end{equation}
 In (\ref{A_tau_BS_bar}), $\bar{\bm{A}}_{\rm BS}^{\mu}\! \in\! \mathbb{C}^{M_{\rm BS}^{\rm h}
 \times L}$ ($\bar{\bm{A}}_{\rm BS}^{\nu}\! \in\! \mathbb{C}^{M_{\rm BS}^{\rm v} \times L}$)
 is the sub-matrix consisting of the first $M_{\rm BS}^{\rm h}$ ($M_{\rm BS}^{\rm v}$) rows
 of $\left(\bm{A}_{\rm BS}^{\mu}\right)^*$ ($\left(\bm{A}_{\rm BS}^{\nu}\right)^*$). In
 this way, $\widetilde{\bm{W}}_u$ visualizes the high-dimensional hybrid analog/digital array
 as a low-dimensional digital array, and $\bm{A}_{\bm{\tau}{\rm BS}}$ holds the triple
 shift-invariance structure for horizontal/vertical AoDs and delays \cite{TSP_Vandermonde98}.
 Therefore, we can apply the MDU-ESPRIT algorithm to obtain the super-resolution estimates
 of horizontal/vertical AoDs and delays, $\big\{ \widehat{\theta}_l^{\rm BS},
 \widehat{\varphi}_l^{\rm BS}, \widehat{\tau}_l \big\}_{l=1}^{\widehat{L}}$.

\subsection{Design Multi-Beam Transmit Precoding Matrix at UD}\label{S3.2}

 We design the transmit precoding matrix $\bm{F}_u\! =\! \bm{F}_{{\rm RF},u}
 \bm{F}_{{\rm BB},u}$ at UD by exploiting the estimate $\big\{
 \widehat{\theta}_l^{\rm UD},\widehat{\varphi}_l^{\rm UD}\big\}_{l=1}^{\widehat{L}}$
 obtained at the downlink CE stage so that the UD with the limited transmit power can
 transmit directional multi-beam signals for improving the received SNR at the BS.

 We first consider the analog transmit precoding matrix $\bm{F}_{{\rm RF},u}\!\in\!
 \mathbb{C}^{N_{\rm UD}\times N_{\rm UD}^{\rm RF}}$.
 The estimate $\widehat{\bm{A}}_{\rm UD}$ of the array response matrix $\bm{A}_{\rm UD}$
 can be calculated given the estimated AoAs $\big\{\widehat{\theta}_l^{\rm UD},
 \widehat{\varphi}_l^{\rm UD}\big\}_{l=1}^{\widehat{L}}$. To fully exploit the acquired
 $\big\{\widehat{\theta}_l^{\rm UD},\widehat{\varphi}_l^{\rm UD}\big\}_{l=1}^{\widehat{L}}$,
 the multi-beam transmit precoding matrix should align its $\widehat{L}$ beams with the
 $\widehat{L}$ estimated AoAs. Specifically, the phase shifters of the PSN at UD can be
 divided into the $\widehat{L}$ groups as equally as possible, depending on $\widehat{L}$
 and $N_{\rm UD}^{\rm RF}$.

\textbf{Case I: $\widehat{L} > N_{\rm UD}^{\rm RF}$}.
 This is the case that the number of beams transmitted by UD is larger than that of RF
 chains. For the UD equipped with the hybrid array with the fully-connected PSN, there are
 $N_{\rm PS}\! =\! N_{{\rm UD}}^{{\rm RF}}N_{{\rm UD}}$ phase shifters. Let the number of
 phase shifters assigned to the $l$th group be $n_{{\rm ps},l}$ with $1\le l \le
 \widehat{L}$. We introduce the $\widehat{L}$-dimensional vector $\bm{v}_{\rm ps}$ as
\begin{equation}\label{n_ph_vec} 
\vspace{-1.0mm}
 \bm{v}_{\rm ps} = \big[ n_{{\rm ps},1} \cdots n_{{\rm ps},\widehat{L}} \big]^{\rm T}
  = n_{\rm ps} \bm{1}_{\widehat{L}} + \big[ \bm{1}_{n_{\rm re}}^{\rm T} ~
  \bm{0}_{\widehat{L} - n_{\rm re}}^{\rm T} \big]^{\rm T} ,
\vspace{-1.0mm}
\end{equation}
 where $n_{\rm ps}\! =\! \lfloor N_{\rm PS}/\widehat{L} \rfloor$ and $n_{\rm re}\! =\!
 \text{mod}\big( N_{\rm PS},\widehat{L}\big)$. By defining the index vector $\bm{p}\!
 =\! \left[ 1 ~ 2 \cdots N_{\rm PS}\right]^{\rm T}$, the ordered index set ${\cal P}_l$
 of the $l$th group can be obtained as ${\cal P}_l\! =\! \bm{p}_{\left[\sum\nolimits_{i=1}^{l-1}
 n_{{\rm ps},i}  + 1:\sum\nolimits_{i=1}^l n_{{\rm ps},i}\right]}$ with $|{\cal P}_l|_c\!
 =\! n_{{\rm ps},l}$. Then, we can define the vector $\bm{f}\! =\! \big[ \bm{f}_1^{\rm T}
 \cdots \bm{f}_{\widehat{L}}^{\rm T} \big]^{\rm T}\! \in\! \mathbb{C}^{N_{\rm PS}}$,
 where $\!\bm{f}_l\! =\! \widehat{\bm{A}}_{{\rm UD}\left[ \text{mod}\left( {\cal P}_l,
 N_{\rm UD} \right),l\right]}^{*}\! \in\! \mathbb{C}^{n_{{\rm ps},l}}$ for $1\! \le\!
 l\! \le\! \widehat{L}$, to obtain $\bm{F}_{{\rm RF},u}\! =\! \text{mat}\left( \bm{f};
 N_{\rm UD},N_{\rm UD}^{\rm RF}\right)$.

\textbf{Case II: $\widehat{L} \le N_{\rm UD}^{\rm RF}$ and $N_{\rm UD}^{\rm RF}$ can be
 divided exactly by $\widehat{L}$}.
 The $N_{\rm UD}^{\rm RF}$ RF chains can be equally allocated to the $\widehat{L}$ groups,
 and we can choose $\bm{F}_{{\rm RF},u}\! =\! \bm{1}_{N_{\rm rep}}^{\rm T} \otimes
 \widehat{\bm{A}}_{\rm UD}^*$, with $N_{\rm rep}\! =\! N_{\rm UD}^{\rm RF}/\widehat{L}$.

\textbf{Case III: $\widehat{L} < N_{\rm UD}^{\rm RF}$ and $N_{\rm UD}^{\rm RF}$ cannot be
 divided exactly by $\widehat{L}$}.
 In this case, we have $\bm{F}_{{\rm RF},u}\! =\! \left[\bm{F}_{{\rm RF},u}^1 ~
 \bm{F}_{{\rm RF},u}^2 \right]$, where $\bm{F}_{{\rm RF},u}^1\! \in\! \mathbb{C}^{N_{\rm UD}
 \times \widehat{L}N_{\rm rep}}$ with $N_{\rm rep}\! =\! \lfloor
 N_{\rm UD}^{\rm RF}/\widehat{L} \rfloor$ and $\bm{F}_{{\rm RF},u}^2\! \in\!
 \mathbb{C}^{N_{\rm UD} \times N_{{\rm UD},{\rm re}}^{\rm RF}}$ with $N_{{\rm UD},{\rm re}}^{\rm RF}
 \! =\! \text{mod}\big(N_{\rm UD}^{\rm RF},\widehat{L}\big)$. We can design
 $\bm{F}_{{\rm RF},u}^1\! =\! \bm{1}_{N_{\rm rep}}^{\rm T}\! \otimes\! \widehat{\bm{A}}_{\rm UD}^*$
 similar to Case II, and we can choose $\bm{F}_{{\rm RF},u}^2\! =\! \text{mat}\big( \big[
 \widetilde{\bm{f}}_1^{\rm T}\cdots \widetilde{\bm{f}}_{\widehat{L}}^{\rm T} \big]^{\rm T};
 N_{\rm UD},N_{{\rm UD},{\rm re}}^{\rm RF} \big)$ similar to Case I, where $\widetilde{\bm{f}}_l$
 can be acquired by using $N_{{\rm UD},{\rm re}}^{\rm RF}$ instead of $N_{\rm UD}^{\rm RF}$ in Case I.

 Due to the limited resolution of the PSN, the phase value of every element in the designed
 $\bm{F}_{{\rm RF},u}$ is quantized to the nearest value in the phase set ${\cal A}$.
 As for the digital transmit precoding matrix $\bm{F}_{{\rm BB},u}\!\in\!
 \mathbb{C}^{N_{\rm UD}^{\rm RF} \times N_{\rm s}^u}$, we can design its element as
 $\left[ \bm{F}_{{\rm BB},u} \right]_{j_1,j_2}\! =\! e^{\textsf{j}2\pi b_{j_1,j_2}}$ with
 $b_{j_1,j_2} \sim {\cal U}[0, ~ 1]$. Finally, we obtain the multi-beam transmit precoding
 matrix $\bm{F}_u\! =\! \bm{F}_{{\rm RF},u}\bm{F}_{{\rm BB},u}$.

 To intuitively compare $\bm{F}_d$ of the BS designed at the downlink CE stage and
 $\bm{F}_u$ of the UD designed at the uplink CE stage, we provide the comparison of
 beam patterns in Fig.~\ref{FIG4}, where we have $N_{{\rm BS}}^{{\rm RF}}\! =\!
 N_{{\rm UD}}^{{\rm RF}}\! =\! 4$ RF chains at the BS and each UD, and the AoAs of the
 $5$ MPCs are known to the UD. Specifically, Fig.~\ref{FIG4}\,(a) depicts the beam
 pattern of the random transmit precoding matrix $\bm{F}_d$ for the $8\! \times\! 8$
 UPA, while Fig.~\ref{FIG4}\,(b) and Fig.~\ref{FIG4}\,(c) plot the beam patterns of
 the multi-beam transmit precoding matrices $\bm{F}_u$ for the $8\! \times\! 8$ and
 $16\! \times\! 16$ UPAs, respectively. Compared with the beam pattern of
 Fig.~\ref{FIG4}\,(a), the beam pattern in Fig.~\ref{FIG4}\,(b) has 5 mainlobes aligned
 with the directions of the AoAs of the 5 MPCs, which can significantly improve the
 SNR at receiver. Moreover, by comparing Fig.~\ref{FIG4}\,(b) with Fig.~\ref{FIG4}\,(c),
 it can be observed that the sidelobes of the multi-beam signals are further suppressed
 when the array size increases. In a nutshell, the proposed multi-beam transmit precoding
 matrix design enables the UD with limited transmit power to form the directional signals
 with multiple beams aligned with the estimated AoAs of the MPCs for improving the
 uplink CE performance.

\section{MDU-ESPRIT Algorithm}\label{S4}

 Because the aggregated receive combining matrix $\widetilde{\bm{W}}_d$ ($\widetilde{\bm{W}}_u$)
 designed at the UD (BS) reconstructs the double (triple) shift-invariance structure of array response,
 spectrum estimation techniques can be utilized to estimate the channel parameters. We consider
 $R$-dimensional ($R$-D) unitary ESPRIT algorithm with $R\!\ge\!2$. Without loss of generality,
 we define a general signal transmission model for the channel consisting of $L$ MPCs and $R$
 sets of spatial frequencies as
\begin{equation}\label{Y_model} 
 \bm{Y} = \bm{A}\bm{S} + \bm{N} ,
\end{equation}
 where $\bm{Y}\! \in\! \mathbb{C}^{M \times N}$ is the received data matrix aggregated over
 $N$ snapshots, $M \!=\! \prod\nolimits_{r=1}^R M_r$, with $M_r$ being the dimension of the
 parameter vector associated with the $r$th spatial frequency for $1\! \le\! r\! \le \! R$,
 and $\bm{S}\! \in\! \mathbb{C}^{L\times N}$ and $\bm{N}\! \in\! \mathbb{C}^{M \times N}$ are
 the transmit signal and noise matrices, respectively, while the array response matrix
 $\bm{A}\! \in\! \mathbb{C}^{M \times L}$ is given by
\begin{equation}\label{A_model} 
\begin{split}
 \bm{A} &= \bm{A}_{\mu_R} \odot \cdots \odot \bm{A}_{\mu_2} \odot \bm{A}_{\mu_1}\\
 &= \left[ \bm{a}\left( \mu_1^1,\mu_1^2,\cdots ,\mu_1^R \right) \cdots \bm{a}\left(
  \mu_L^1,\mu_L^2,\cdots ,\mu_L^R \right) \right] .
\end{split}
\end{equation}
 In (\ref{A_model}), $\bm{A}_{\mu_r}\! =\! \left[ \bm{a}(\mu_1^r)\cdots \bm{a}(\mu_L^r)
 \right]\! \in\! \mathbb{C}^{M_r \times L}$ is the steering matrix related to the $r$th
 set of spatial frequencies $\{\mu_l^r\}_{l=1}^L$, with $\bm{a}\left(\mu_l^r\right)\!
 =\! \left[ 1 ~ e^{\textsf{j}\mu_l^r}\cdots e^{\textsf{j}(M_r-1)\mu_l^r} \right]^{\rm T}
 \! \in\! \mathbb{C}^{M_r}$ being the $l$th steering vector, while the array response
 vector $\bm{a}\left( \mu_l^1,\mu _l^2, \cdots, \mu_l^R\right)\! \in\! \mathbb{C}^M$
 related to the $l$th MPC is given by
\begin{equation}\label{Steering_Vector}  
 \bm{a}\left( \mu_l^1,\mu_l^2,\cdots ,\mu_l^R\right) = \bm{a}\left( \mu_l^R \right)
  \otimes \cdots \otimes \bm{a}\left( \mu_l^2 \right) \otimes \bm{a}\left( \mu_l^1 \right) .
\end{equation}

 The MDU-ESPRIT algorithm, which acquires the super-resolution estimates of the $R$
 sets of spatial frequencies from (\ref{Y_model}), denoted by $\left\{\widehat{\mu}_l^r
 \right\}_{l=1}^L$ for $1\! \le\! r\! \le\! R$,  consists of the five steps.

\subsubsection{$R$-D Spatial Smoothing Preprocessing (SSP)} 

 In order to take into account the insufficient measurement dimension $N$ caused by the
 limited training overhead, we will exploit the spatial smoothing technique \cite{TSP_JADE98}
 to preprocess the original data matrix $\bm{Y}$ of (\ref{Y_model}). This preprocessing
 can mitigate the influence of other coherent signals and avoid the rank deficiency of the
 covariance matrix of $\bm{S}$ to enhance robustness. Specifically, we first define the
 $R$ spatial smoothing parameters $\{ G_r \}_{r=1}^R$ with $1\! \le\! G_r\! \le\! M_r$,
 and obtain the sub-dimensions $\{ M_r^{\rm sub} \}_{r=1}^R$ corresponding to
 $\{ M_r \}_{r=1}^R$ as $M_r^{ \rm sub}\! =\! M_r\! -\! G_r\! +\! 1$ for
 $1\! \le\! r\! \le\! R$. Thus the size of total sub-dimension is $M_{\rm sub}\! =\!
 \prod\nolimits_{r=1}^R M_r^{\rm sub}$. To obtain the $R$-D selection matrix, we next
 define the $g_r$th `1-D' selection matrix as $\bm{J}^{(g_r)}\! =\! \left[\bm{O}_{M_r^{\rm sub}
 \times (g_r-1)} ~ \bm{I}_{M_r^{\rm sub}} ~ \bm{O}_{M_r^{\rm sub}\times (G_r-g_r)} \right]\!
 \in\! \mathbb{R}^{M_r^{\rm sub} \times M_r}$ for $1\!\le\! g_r\!\le\! G_r$. Then, we
 can obtain $G\! =\! \prod\nolimits_{r=1}^R G_r$ `$R$-D' selection matrices, with the
 $(g_1,g_2,\cdots ,g_R)$th `$R$-D' selection matrix given by $\bm{J}_{g_1,g_2,\cdots ,g_R}
 \! =\! \bm{J}^{(g_R)} \otimes  \cdots  \otimes \bm{J}^{(g_2)} \otimes \bm{J}^{(g_1)}
 \! \in\! \mathbb{R}^{M_{\rm sub} \times M}$. By applying these $R$-D selection matrices
 to $\bm{Y}$, the smoothed complex-valued data matrix $\bar{\bm{Y}}\! \in\!
 \mathbb{C}^{M_{\rm sub} \times NG}$ is obtained as
\begin{equation}\label{Y_SS}  
\begin{split}
 \bar{\bm{Y}} = &\big[ \left(\bm{J}_{1,1,\cdots\!,1,1}\bm{Y}\right) \, \cdots\,
  \left(\bm{J}_{1,1,\cdots\!,1,G_R}\bm{Y}\right) \, \left(\bm{J}_{1,1,\cdots\!,2,1}\bm{Y}\right)\\
  &\left(\bm{J}_{1,1,\cdots\!,2,2}\bm{Y}\right)\, \cdots\, \left(\bm{J}_{G_1,G_2, \cdots\!,G_{R-1},G_R}\bm{Y}\right) \big] .
\end{split}
\end{equation}

\subsubsection{Real-Valued Processing (RVP)} 

 To reduce the computational complexity, the forward backward averaging technique
 \cite{{2D_U_ESPRIT96}} is utilized to transform $\bar{\bm{Y}}$ into the real-valued
 matrix $\bar{\bm{Y}}_{\rm re}\! \in\! \mathbb{R}^{M_{\rm sub} \times 2NG}$
\begin{equation}\label{H_SS_Re} 
 \bar{\bm{Y}}_{\rm re} = \bm{Q}_{M_{\rm sub}}^{\rm H}\left[\bar{\bm{Y}} ~ \left(
  \bm{\varPi}_{M_{\rm sub}}\bar{\bm{Y}}^* \bm{\varPi}_{NG}\right) \right] \bm{Q}_{2NG} ,
\end{equation}
 where $\bm{\varPi}_n$ is the exchange matrix of size $n \times n$ that permutates the
 row order of $\bm{I}_n$, and $\bm{Q}_n\! \in\! \mathbb{C}^{n \times n}$ is a sparse
 unitary matrix satisfying $\bm{\varPi}_n\bm{Q}_n^{\rm H}\! =\! \bm{Q}_n$.

\subsubsection{Signal Subspace Approximation (SSA)} 

 To extract the information of spatial frequencies from the real-valued matrix
 $\bar{\bm{Y}}_{\rm re}$, we introduce the transform steering matrix $\bm{K}$
  which satisfies \cite{{2D_U_ESPRIT96}}
\begin{equation}\label{transformed_equations} 
 \Re\left\{ \bm{Q}_{m_r}^{\rm H} \bm{J}_r \bm{Q}_{M_{\rm sub}}\right\} \bm{K} \bm{\varLambda}_r =
 \Im\left\{ \bm{Q}_{m_r}^{\rm H} \bm{J}_r \bm{Q}_{M_{\rm sub}}\right\} \bm{K} ,
\end{equation}
 where $1 \le r \le R$, $m_r\! =\! M_{\rm sub} \left(M_r^{\rm sub}\! -\! 1\right)/M_r^{\rm sub}$, and
 $\bm{\varLambda}_r\! =\! \text{diag}\big( \left[\tan\left( \mu_1^r/2 \right) \cdots
 \tan\left( \mu_L^r/2\right)\right]^{\rm T} \big)$ is the real-valued diagonal matrix
 involving the desired spatial frequencies $\{ \mu_l^r \}_{l=1}^L$, while
 $\bm{J}_r\! =\! \bm{I}_{\prod\nolimits_{i=r+1}^R M_i^{\rm sub}}\! \otimes\!
 \widetilde{\bm{J}}^{(r)}\! \otimes\! \bm{I}_{\prod\nolimits_{i=1}^{r-1} M_i^{\rm sub}}\!
 \in\! \mathbb{R}^{m_r \times M_{\rm sub}}$, with $\widetilde{\bm{J}}^{(r)}\! =\!
 \left[ \bm{0}_{M_r^{\rm sub} - 1} ~ \bm{I}_{M_r^{\rm sub} - 1} \right]$.
 Note that $\bm{K}$ is related to the approximate signal subspace matrix $\bm{E}_{\rm s}
 \! \in\! \mathbb{R}^{M_{\rm sub} \times L}$ corresponding to the underlying signal
 subspace. Specifically, since the columns of $\bm{K}$ and $\bm{E}_{\rm s}$ span the same
 $L$-dimensional signal subspace \cite{TSP_JADE98,TSP_Vandermonde98},
 $\bm{K}\! =\! \bm{E}_{\rm s} \bm{T}$, where $\bm{T}\! \in\! \mathbb{R}^{L \times L}$
 is a non-singular matrix. To determine $\bm{E}_{\rm s}$, we take the left singular
 vectors corresponding to the largest $L$ singular values of $\bar{\bm{Y}}_{\rm re}$
 as $\bm{E}_{\rm s}$. Specifically, from the real-valued partial singular values
 decomposition (SVD) of $\bar{\bm{Y}}_{\rm re}\! =\! \bm{U}_{\rm re} \bm{\varSigma}_{\rm re}
 \bm{V}_{\rm re}^{\rm H}$, we have $\bm{E}_{\rm s}\! =\! \bm{U}_{{\rm re}[:,1:L]}$.

\subsubsection{Shift-Invariance Equation Solving (SIES)} 

 Based on the acquired approximate signal subspace $\bm{E}_{\rm s}$, we use $\bm{K}\!
 =\! \bm{E}_{\rm s}\bm{T}$ in (\ref{transformed_equations}) to obtain the $R$
 shift-invariance equations
\begin{equation}\label{shift_invariance_equations} 
 \Re \left\{ \bm{Q}_{m_r}^{\rm H} \bm{J}_r \bm{Q}_{M_{\rm sub}} \right\} \bm{E}_{\rm s}
  \bm{\varPhi}_r = \Im \left\{ \bm{Q}_{m_r}^{\rm H} \bm{J}_r \bm{Q}_{M_{\rm sub}} \right\}
  \bm{E}_{\rm s} ,
\end{equation}
 where $\bm{\varPhi}_r\! =\! \bm{T} \bm{\varLambda}_r \bm{T}^{-1}\! \in\! \mathbb{R}^{L \times L}$
 for $1 \le r \le R$. To estimate the diagonal matrices $\{ \bm{\varLambda}_r \}_{r=1}^R$,
 we first obtain the estimates of the $R$ real-valued matrices $\{\bm{\varPhi}_r \}_{r=1}^R$,
 denoted as $\big\{ \widehat{\bm{\varPhi}}_r\big\}_{r=1}^R$, by applying the
 LS or total least squares (TLS) estimator to solve the $R$ shift-invariance
 equations of (\ref{shift_invariance_equations}).

\subsubsection{$R$-D Joint Diagonalization (JD)} 

 From the estimated $\widehat{\bm{\varPhi}}_r\! =\! \bm{T}\widehat{\bm{\varLambda}}_r
 \bm{T}^{-1}$ with $\widehat{\bm{\varLambda}}_r$ denoting the estimate of $\bm{\varLambda}_r$
 for $1\! \le\! r\! \le\! R$, we exploit the following $R$-D joint diagonalization to
 obtain the paired estimates of the spatial frequencies $\left\{ \widehat{\mu}_l^r
 \right\}_{l=1}^L$ from $\big\{ \widehat{\bm{\varLambda}}_r \big\}_{r=1}^R$.
 Specifically, we consider the two cases of $R\! =\! 2$ and $R\! \ge\! 3$. For $R\! =\! 2$,
 namely, the 2-D case, since $\widehat{\bm{\varPhi}}_1$ and $\widehat{\bm{\varPhi}}_2$
 share the same eigenvector matrix $\bm{T}$, we can calculate the EVD of the complex-valued
 matrix $\bm{\varPsi}\! =\! \widehat{\bm{\varPhi}}_1\! +\! \textsf{j}\widehat{\bm{\varPhi}}_2$
 to obtain $\widehat{\bm{\varLambda}}_1$ and $\widehat{\bm{\varLambda}}_2$, specifically,
 $\bm{\varPsi}\! =\! \bm{T}\bm{\varDelta}\bm{T}^{-1}$ with $\bm{\varDelta}\! =\!
 \widehat{\bm{\varLambda}}_1\! +\! \textsf{j}\widehat{\bm{\varLambda}}_2$, and
 $\widehat{\bm{\varLambda}}_1\! =\! \Re \{ \bm{\varDelta} \}$ and $\widehat{\bm{\varLambda}}_2
 \! =\! \Im\{ \bm{\varDelta} \}$. For $R\! \ge\! 3$, the noise-corrupted matrices
 $\big\{ \widehat{\bm{\varPhi}}_r \big\}_{r=1}^R$ do not always exactly share the
 same $\bm{T}$. Hence, we exploit the simultaneous Schur decomposition (SSD) algorithm
 \cite{{SSD_R_U_ESPRIT98}}, which is developed from the real Schur decomposition
 \cite{{Matrix13}} for multi-parameter estimation and pairing. By utilizing the SSD
 algorithm, we obtain the $R$ approximate upper-triangular matrices $\left\{
 \bm{\varGamma}_r\right\}_{r=1}^R$ so that $\big\{ \widehat{\bm{\varLambda}}_r \big\}_{r=1}^R$
 are acquired as the main diagonal elements of $\left\{ \bm{\varGamma}_r\right\}_{r=1}^R$,
 i.e., $\widehat{\bm{\varLambda}}_r
 \! =\! \text{diag}\left(\text{vdiag}( \bm{\varGamma}_r)\right)$, for $1\!\le\! r\!\le\! R$.
 Finally, the $R$ paired super-resolution estimates of the spatial frequencies $\left\{
 \widehat{\mu}_l^r\right\}_{l=1}^L$ can be calculated from $\widehat{\bm{\varLambda}}_r$
 as $\widehat{\mu}_l^r\! =\! 2\arctan \big(\big[\widehat{\bm{\varLambda}}_r \big]_{l,l}
 \big)$ for $1\! \le\! l\! \le\! L$ and $1\! \le\! r\! \le\! R$.

 This MDU-ESPRIT algorithm is summarized in Algorithm~\ref{ALG2}. At the downlink CE
 stage, we estimate $\left\{ \widehat{\mu}_l^{\rm UD},\widehat{\nu}_l^{\rm UD}
 \right\}_{l=1}^{\widehat L}$ based on $\bar{\bm{Y}}_d$ of (\ref{Y_bar_D}) by applying
 the 2-D ($R\! =\! 2$) unitary ESPRIT algorithm. Furthermore, based on $\bar{\bm{Y}}_u$
 of (\ref{Y_bar_U}), $\left\{\widehat{\mu}_l^{\rm BS}, \widehat{\nu}_l^{\rm BS},
 \widehat{\mu}_l^{\tau} \right\}_{l=1}^{\widehat{L}}$ are estimated using the 3-D
 ($R\! =\! 3$) unitary ESPRIT algorithm at the uplink CE stage. Hence, the spatial
 smoothing parameters for Algorithm~\ref{ALG2} in the 2-D and 3-D cases are $\{G_1^d,G_2^d\}$
 and $\{G_1^u,G_2^u,G_3^u\}$, respectively. The corresponding total sub-dimension sizes at
 the UD and BS are $M_{\rm sub}^{\rm UD}\! =\! (M_{\rm UD}^{\rm h}\! -\! G_1^d\! +\! 1)
 (M_{\rm UD}^{\rm v}\! -\!G_2^d\! +\! 1)$ and $M_{\rm sub}^{\rm BS}\! =\!
 (M_{\rm BS}^{\rm h}\! -\! G_1^u\! +\! 1)(M_{\rm BS}^{\rm v}\! -\!G_2^u\! +\! 1)
 (K\! -\! G_3^u\! +\! 1)$, respectively.

\begin{algorithm}[tp!]
\caption{MDU-ESPRIT Algorithm}
\label{ALG2}
\begin{algorithmic}[1]
\REQUIRE Data matrix $\bm{Y}$, number of MPCs $L$, sub-dimensions $\{ M_r \}_{r=1}^R$,
  spatial smoothing parameters $\{ G_r \}_{r=1}^R$
\ENSURE Super-resolution estimates of spatial frequencies $\left\{ \widehat{\mu}_l^r
  \right\}_{l=1}^L$, $1\le r\le R$
\STATE Obtain smoothed data matrix $\bar{\bm{Y}}$ (\ref{Y_SS}) using $R$-D spatial
  smoothing preprocessing
\STATE Obtain real-valued data matrix $\bar{\bm{Y}}_{\rm re}$ (\ref{H_SS_Re}) using
  forward backward averaging
\STATE Determine approximate signal subspace matrix $\bm{E}_{\rm s}$ through SVD
\STATE Solve shift-invariance equations (\ref{shift_invariance_equations}) to obtain
  $R$ real-valued matrices $\big\{ \widehat{\bm{\varPhi}}_r \big\}_{r=1}^R$
\STATE Perform $R$-D joint diagonalization to estimate diagonal matrices $\big\{
  \widehat{\bm{\varLambda}}_r\big\}_{r=1}^R$: i)~$R = 2$, calculate EVD of
  $\bm{\varPsi}=\widehat{\bm{\varPhi}}_1 + \textsf{j}\widehat{\bm{\varPhi}}_2 =
  \bm{T}\bm{\varDelta}\bm{T}^{-1}$ to obtain $\widehat{\bm{\varLambda}}_1 =
  \Re \{ \bm{\varDelta}\}$ and $\widehat{\bm{\varLambda}}_2 = \Im \{ \bm{\varDelta}\}$;
  ii)~$R \ge 3$, obtain $R$ diagonal matrices $\big\{ \widehat {\bm{\varLambda}}_r
  \big\}_{r=1}^R$ via SSD algorithm
\STATE Extract $R$ paired $\left\{ \widehat{\mu}_l^r \right\}_{l=1}^L$ from
   $\big\{ \widehat{\bm{\varLambda}}_r \big\}_{r=1}^R$
\end{algorithmic}
\end{algorithm}

\section{ML Pairing and Path Gains Estimation}\label{S5}

 At the downlink CE stage, the BS obtains the estimated horizontal/vertical AoAs
 $\left\{\bar{\theta}_l^{\rm UD},\bar{\varphi}_l^{\rm UD} \right\}_{l=1}^{\widehat{L}}$
 fed back by the UD. It then estimates the horizontal/vertical AoDs and delays
 $\big\{ \widehat{\theta}_l^{\rm BS},\!\widehat{\varphi}_l^{\rm BS},\!
 \widehat{\tau}_l\big\}_{l=1}^{\widehat{L}}$ at the uplink CE stage. Since
 $\left\{\bar{\theta}_l^{\rm UD},\!\bar{\varphi}_l^{\rm UD} \right\}_{l=1}^{\widehat{L}}$
 and $\big\{ \widehat{\theta}_l^{\rm BS},\!\widehat{\varphi}_l^{\rm BS},\!\widehat{\tau}_l
 \big\}_{l=1}^{\widehat{L}}$ are acquired in the two different ends of the channel at
 two different stages, it is necessary to pair them.  Furthermore, the path gain vector
 $\bm{\alpha}$ needs to be estimated. We propose to apply an ML approach to pair
 the channel parameters and to estimate the path gains, which corresponds to Step 7 of
 Fig.~\ref{FIG2} at the BS.

 Specifically, according to (\ref{A_tau_BS_bar}), we construct the equivalent steering
 matrix associated with $\big\{ \widehat{\theta}_l^{\rm BS},\widehat{\varphi}_l^{\rm BS},
 \widehat{\tau}_l\big\}_{l=1}^{\widehat{L}}$ as $\widehat{\bm{A}}_{\tau{\rm BS}}$,
 where $\{ \widehat{\tau}_l\}_{l=1}^{\widehat{L}}$ are arranged in ascending order.
 Based on the estimated horizontal/vertical AoAs fed back to the BS,
 $\left\{\bar{\theta}_l^{\rm UD},\bar{\varphi}_l^{\rm UD} \right\}_{l=1}^{\widehat{L}}$,
 we can reconstruct the estimated multi-beam transmit precoding matrix, denoted by
 $\widehat{\bm{F}}_u$, similar to the construction of $\bm{F}_u$ given in Section~\ref{S3.2}.
 Clearly, there are a total of $J_{\rm c}\! =\! \widehat{L} !$ possible ordered
 combinations or pairs $\big\{\bar{\theta}_{l_j}^{\rm UD},\bar{\varphi}_{l_j}^{\rm UD}
 \big\}_{l_j=1}^{\widehat{L}}$ that can pair with $\widehat{\bm{A}}_{\tau{\rm BS}}$ or
 $\big\{ \widehat{\theta}_l^{\rm BS},\widehat{\varphi}_l^{\rm BS}, \widehat{\tau}_l
 \big\}_{l=1}^{\widehat{L}}$, where $j\! \in\! {\cal J}$ and the size of the ordered
 set ${\cal J}$ is $|{\cal J}|_c\! =\! J_{\rm c}$. For each $\big\{
 \bar{\theta}_{l_j}^{\rm UD},\bar{\varphi}_{l_j}^{\rm UD}\big\}_{l_j=1}^{\widehat{L}}$,
 we can establish the corresponding array response matrix $\bm{A}_{\rm UD}$, which  is
 denoted as $\widehat{\bm{A}}_{{\rm UD},j}$. Thus, for each pair of the AoDs and AoAs, we have
 $\widehat{\bm{A}}_{\tau {\rm BS}}$, $\widehat{\bm{A}}_{{\rm UD},j}$ and $\widehat{\bm{F}}_u$.
 Substituting them into (\ref{y_U_tilde}) yields $\check{\bm{y}}_u\! =\! \widehat{\bm{A}}_j
 \bm{\alpha}_j\! + \check{\bm{n}}_u$, where $\widehat{\bm{A}}_j\! =\! \widehat{\bm{A}}_{\tau {\rm BS}}
 \! \odot\! \big( \widehat{\bm{A}}_{{\rm UD},j}^{\rm T} \widehat{\bm{F}}_u \bm{S}_u
 \big)^{\rm T}$ while $\bm{\alpha}_j$ is the path gain vector corresponding to the $j$th
 pair of the AoDs and AoAs with $j\! \in\! {\cal J}$. The LS estimate of $\bm{\alpha}_j$
 is readily given as
\begin{equation}\label{Alpha_j_hat} 
 \widehat{\bm{\alpha}}_j = \big( \widehat{\bm{A}}_j^{\rm H}\widehat{\bm{A}}_j \big)^{-1}
   \widehat{\bm{A}}_j^{\rm H} \check{\bm{y}}_u .
\end{equation}
 From the estimate $\widehat{\bm{\alpha}}_j$, we can estimate $\check{\bm{y}}_u$ according
 to $\widehat{\check{\bm{y}}}_{u,j}\! =\! \widehat{\bm{A}}_j\widehat{\bm{\alpha}}_j$ with
 the  residual $\big\| \check{\bm{y}}_u\! -\! \widehat{\check{\bm{y}}}_{u,j}\big\|_2^2$.
 We can then find the optimal pair index $j^{\star}$ by solving the following optimization
 problem
\begin{equation} \label{Optimization_ML} 
 j^{\star} = \arg \min\limits_{j\in {\cal J}} \big\| \check{\bm{y}}_u -
  \widehat{\check{\bm{y}}}_{u,j}\big\|_2^2 .
\end{equation}
 Hence, the optimal estimate of the path gains is given by $\widehat{\bm{\alpha}}\!
 =\! \widehat{\bm{\alpha}}_{j^{\star}}\! =\! \beta \left[ \widehat{\alpha}_1 \cdots
 \widehat{\alpha}_{\widehat{L}} \right]^{\rm T}$ and we have the optimal ordered mmWave
 channel parameter estimate $\big\{ \bar{\theta}_l^{\rm UD},\bar{\varphi}_l^{\rm UD},
 \widehat{\theta}_l^{\rm BS},\widehat{\varphi}_l^{\rm BS},\widehat{\tau}_l,
 \widehat{\alpha}_l\big\}_{l=1}^{\widehat{L}}$.

 By substituting $\big\{ \bar{\theta}_l^{\rm UD},\bar{\varphi}_l^{\rm UD},
 \widehat{\theta}_l^{\rm BS},\widehat{\varphi}_l^{\rm BS},\widehat{\tau}_l,
 \widehat{\alpha}_l\big\}_{l=1}^{\widehat{L}}$ into (\ref{frequency_domain_channel}), we
 obtain the optimally estimated frequency-domain channel matrix $\widehat{\bm{ H}}[k]$ at
 the $k$th subcarrier as
\begin{equation}\label{frequency_domain_channel_estimation} 
 \widehat{\bm{H}}[k]\! =\! \beta \sum\limits_{l=1}^{\widehat{L}} \widehat{\alpha}_{l}
  \bm{a}_{\rm UD}\left( \bar{\mu} _l^{\rm UD},\bar{\nu}_l^{\rm UD} \right)
  \bm{a}_{\rm BS}^{\rm H} \left( \widehat{\mu}_l^{\rm BS},\widehat{\nu}_l^{\rm BS} \right)
  e^{-\textsf{j}2\pi \frac{k f_s}{K} \widehat{\tau}_l} ,
\end{equation}
 where $\bar{\mu}_l^{\rm UD}\! =\! \pi\sin (\bar{\theta}_l^{\rm UD})\cos (\bar{\varphi}_l^{\rm UD})$
 and $\bar{\nu}_l^{\rm UD}\! =\! \pi\sin (\bar{\varphi}_l^{\rm UD})$, while $\widehat{\mu}_l^{\rm BS}
 \! =\! \pi\sin\big(\widehat{\theta}_l^{\rm BS}\big)\cos\left(\widehat{\varphi}_l^{\rm BS}\right)$
 and $\widehat{\nu}_l^{\rm BS}\! =\! \pi\sin\left(\widehat{\varphi}_l^{\rm BS}\right)$.

\begin{figure*}[!tp]
\captionsetup{font={footnotesize}, name = {Fig.}, labelsep = period} 
\captionsetup[subfigure]{singlelinecheck = on, justification = raggedright, font={footnotesize}}
\centering
\subfloat{
\label{FIG5(a)}
\begin{minipage}[t]{0.4\linewidth}
\centering
\includegraphics[width=2.6in]{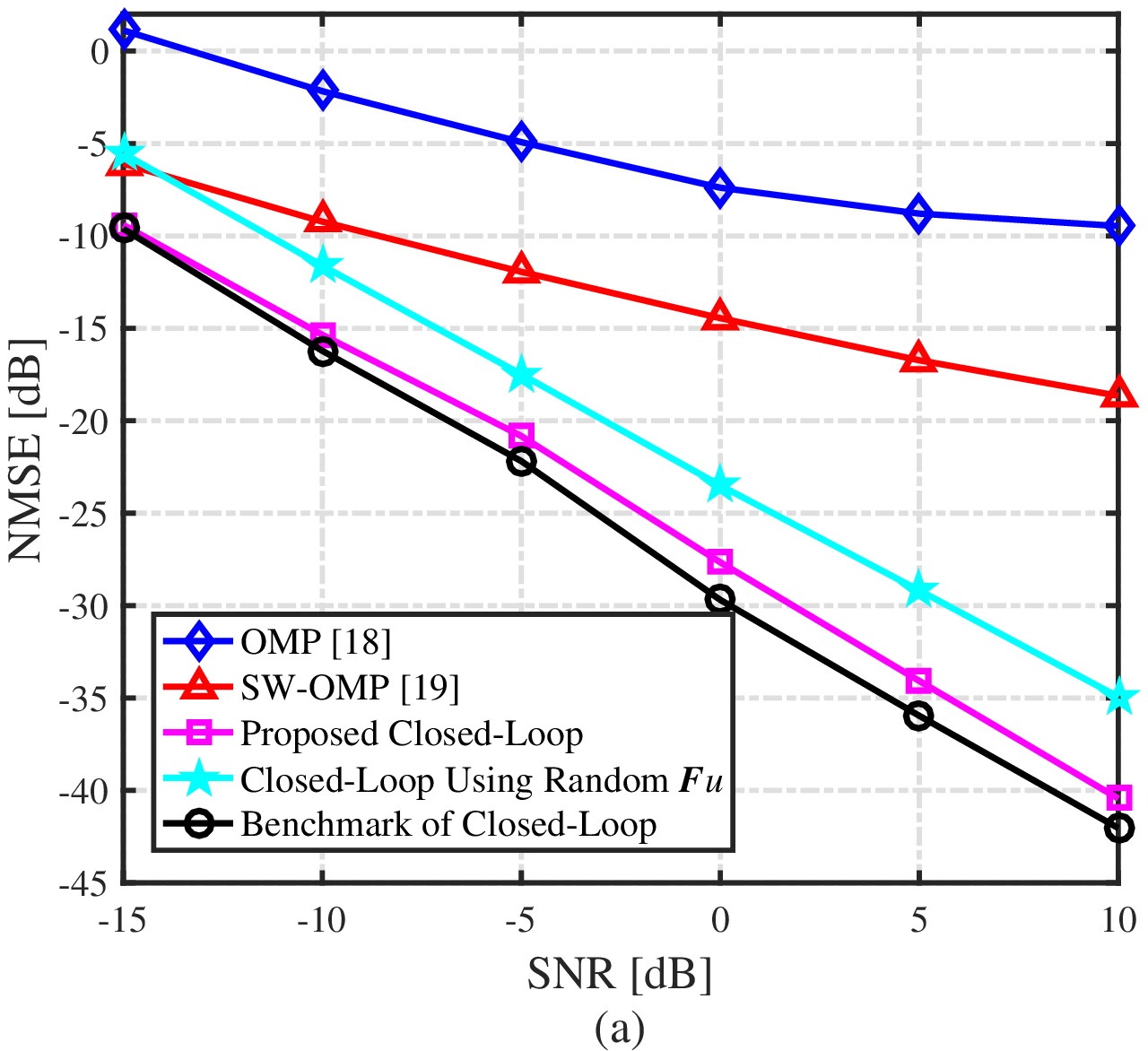}
\end{minipage}
}
\subfloat{
\label{FIG5(b)}
\begin{minipage}[t]{0.4\linewidth}
\centering
\includegraphics[width = 2.6in]{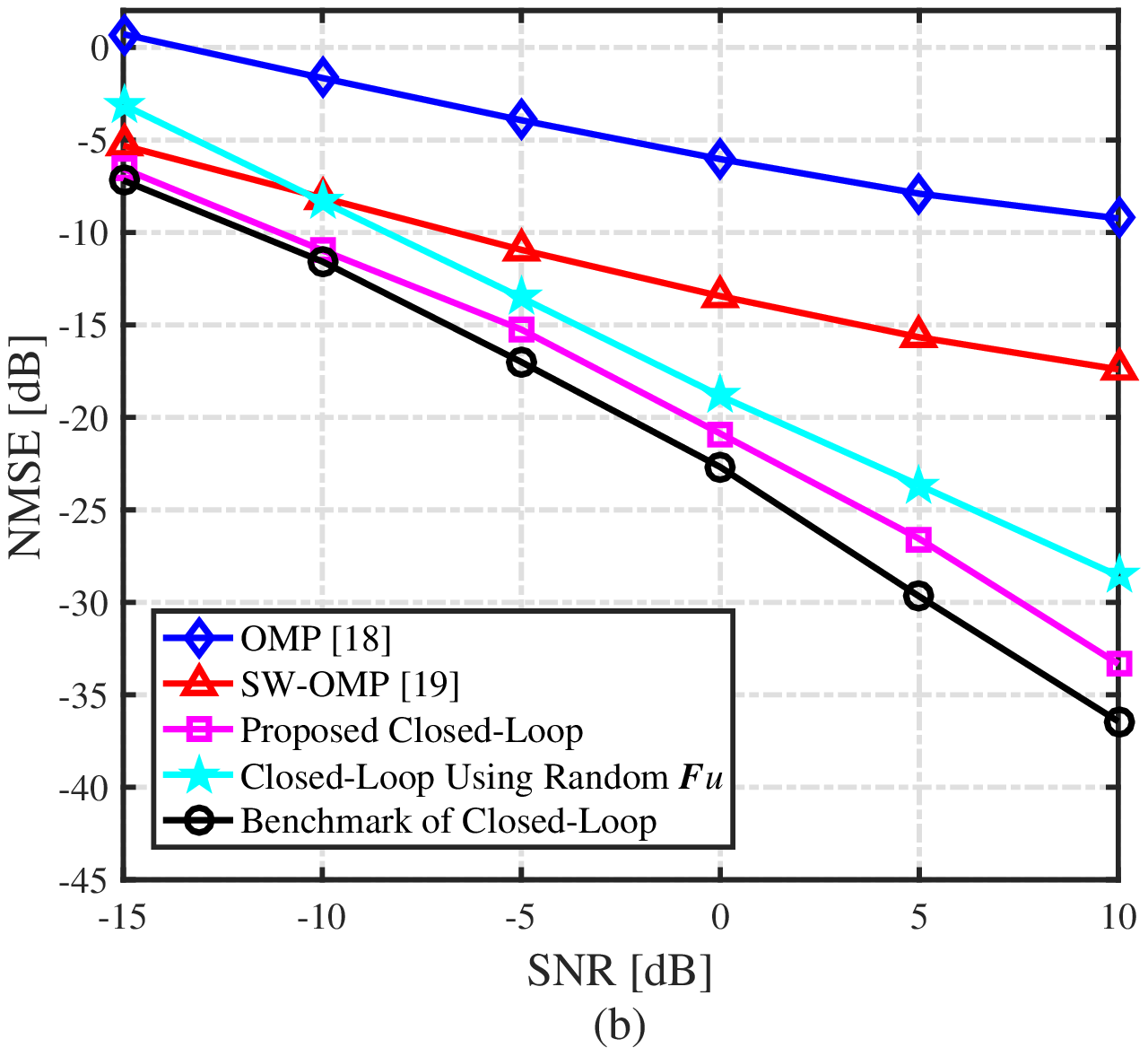}
\end{minipage}
}
\caption{NMSE performance comparison of different CE schemes versus SNRs with the
 same training overhead $T_{\rm CE}=132$: (a)~the number of MPCs $L=3$; and (b)~$L=5$.}
\label{FIG5}
\end{figure*}

\section{Performance Evaluation}\label{S6}

 An extensive simulation investigation is carried out to evaluate the CE performance
 and computational complexity of the proposed closed-loop CE scheme. In simulations,
 the carrier frequency is $f_c\! =\! 30$\,GHz with the bandwidth $f_s\! = \! 200$\,MHz,
 the  numbers of RF chains at BS and UD are both 4, i.e., $N_{\rm BS}^{\rm RF}\! = \!
 N_{\rm UD}^{\rm RF} \! = \! 4$, and the numbers of horizontal and vertical antennas
 at BS and UD are both 12, i.e., $N_{\rm BS}^{\rm h}\! =\! N_{\rm BS}^{\rm v}\! =\!
 N_{\rm UD}^{\rm h}\! =\! N_{\rm UD}^{\rm v}\! =\! 12$, and the quantization accuracy
 of the PSN is defined by $N_q^{\rm ps}\! =\! 3$ bits, while the feedback quantization
 accuracy for AoAs is specified by $N_q^{\rm ang}\! =\! 10$ bits. Without loss
 of generality, the case of single UD $Q\! =\! 1$ is considered. From Fig.~\ref{FIG3},
 it is clear that for the generic case of $Q > 1$, the uplink training overhead becomes
 $N_u N_{\rm o}^u$ instead of $Q N_u N_{\rm o}^u$ for the case of $Q=1$. The channel
 model is simulated as follows. Each of the path gains $\{\alpha_l\}_{l=1}^L$ is
 generated according to ${\cal CN}(0,1)$, while the other channel parameters $\{\tau_l,
 \theta_l^{\rm UD},\theta_l^{\rm BS},\varphi_l^{\rm UD},\varphi_l^{\rm BS}\}_{l=1}^L$
 all follow uniform distribution, specifically, $\tau_l\!\sim\! {\cal U}[0, ~ \tau_{\max}]$
 and $\theta_l^{\rm UD},\theta_l^{\rm BS},\varphi_l^{\rm UD},\varphi_l^{\rm BS}\!\sim\!
 {\cal U}\left[ -\pi /3 ~\pi /3 \right]$ for the $l$th MPC. The maximum multipath delay
 is set to $\tau_{\max}\! =\! 16 T_s$, i.e., $N_{\rm c}\! =\! 16$. The number of subcarriers
 is set to $K\! =\! 128$ with the length of cyclic prefix (CP) being 32, and perfect frame
 synchronization is assumed. In our proposed solution, the sizes of low-dimensional digital
 sub-arrays visualized from the high-dimensional hybrid arrays are set to $M_{\rm BS}^{\rm h}\!
 =\! M_{\rm BS}^{\rm v}\! =\! M_{\rm UD}^{\rm h}\! =\! M_{\rm UD}^{\rm v}\! =\! 8$. Hence,
 the numbers of downlink and uplink training time slots are $N_d\! =\! 22$ and
 $N_u\! =\! 22$, respectively, given the number of downlink independent signal streams
 $N_{\rm s}^d\! = \! N_{\rm UD}^{\rm RF}\! -\! 1\! = \! 3$ and the number of uplink
 independent signal streams $N_{\rm s}^u \! =  \! N_{\rm BS}^{\rm RF}\! -\! 1\!
 = \! 3$. Additionally, $P\! =\! 8$ adjacent subcarriers are jointly employed to estimate
 the number of MPCs, with the threshold parameter $\varepsilon$ empirically set to 1.54,
 0.50, 0.16, 0.05, 0.016, and 0.005, respectively, at the SNR of -15\,dB, -10\,dB, -5\,dB,
 0\,dB, 5\,dB, and 10\,dB. The spatial smoothing parameters used for Algorithm~\ref{ALG2}
 are $G_1^d\! =\! G_2^d\! =\! G_1^u\! =\! G_2^u\! =\! 2$ and $G_3^u \! =\! K/2$. The
 downlink and uplink SNRs are both defined as $\rho\sigma_\alpha^2/\sigma_n^2$, where
 $\rho$ and $\sigma_n^2$ are the transmit power and receiver noise variance, respectively.

 The state-of-the-art OMP-based frequency-domain scheme \cite{JSAC_OMP17}\footnote{The
 redundant dictionary of OMP-based time-domain method in \cite{JSAC_OMP17} is generated
 by the quantized grids at the delay and angle domains, which imposes the unaffordable
 computational complexity and storage requirements. Hence, we just consider the
 frequency-domain scheme in simulations.} and the SW-OMP-based scheme \cite{TWC_SW_OMP18}
 are adopted as two benchmarks. In order to be consistent with \cite{JSAC_OMP17} and
 \cite{TWC_SW_OMP18}, their digital transmit precoding/receive combining matrices are
 taken as the identity matrix, while the design of analog counterparts is similar to
 the construction of $\bm{F}_{{\rm RF},d}[m]$ given in Section II-B. The sizes of the
 quantized angle-domain grids associated with horizontal/vertical AoAs/AoDs, denoted by
 $G_{\rm BS}^{\rm h}$, $G_{\rm BS}^{\rm v}$, $G_{\rm UD}^{\rm h}$ and $G_{\rm UD}^{\rm v}$,
 are set to twice the numbers of antennas in the horizontal and vertical directions of UPA,
 respectively, according to \cite{JSAC_OMP17,TWC_SW_OMP18}, i.e., $G_{\rm BS}\! = \!
 G_{\rm BS}^{\rm h} \times G_{\rm BS}^{\rm v}\! =\! 2 N_{\rm BS}^{\rm h} \times
 2 N_{\rm BS}^{\rm v}\! =\! 24\times 24$ and $G_{\rm UD}\! =\! G_{\rm UD}^{\rm h}\times
 G_{\rm UD}^{\rm v}\! =\! 2 N_{\rm UD}^{\rm h} \times 2 N_{\rm UD}^{\rm v}\! =\! 24 \times 24$.
 Furthermore, all CE schemes adopt the same training overhead, which is equal to the
 required number of training frames \cite{JSAC_OMP17,TWC_SW_OMP18}, to ensure the fairness
 of comparison.

\subsection{CE Performance Evaluation}\label{S6.1}

 First the CE performance is evaluated using the normalized mean square error (NMSE) metric
 given by
\begin{equation}\label{NMSE} 
 \text{NMSE} =  \mathbb{E}\bigg( \sum\limits_{k=0}^{K-1}\big\| \bm{H}[k] - \widehat{\bm{H}}[k] \big\|_F^2
  \Big/\sum\limits_{k=0}^{K-1}\big\| \bm{H}[k]\big\|_F^2 \bigg).
\end{equation}

\begin{figure*}[!tp]
\captionsetup{font={footnotesize}, name = {Fig.}, labelsep = period} 
\captionsetup[subfigure]{singlelinecheck = on, justification = raggedright, font={footnotesize}}
\centering
\subfloat{
\label{FIG6(a)}
\begin{minipage}[t]{0.4\linewidth}
\centering
\includegraphics[width = 2.6in]{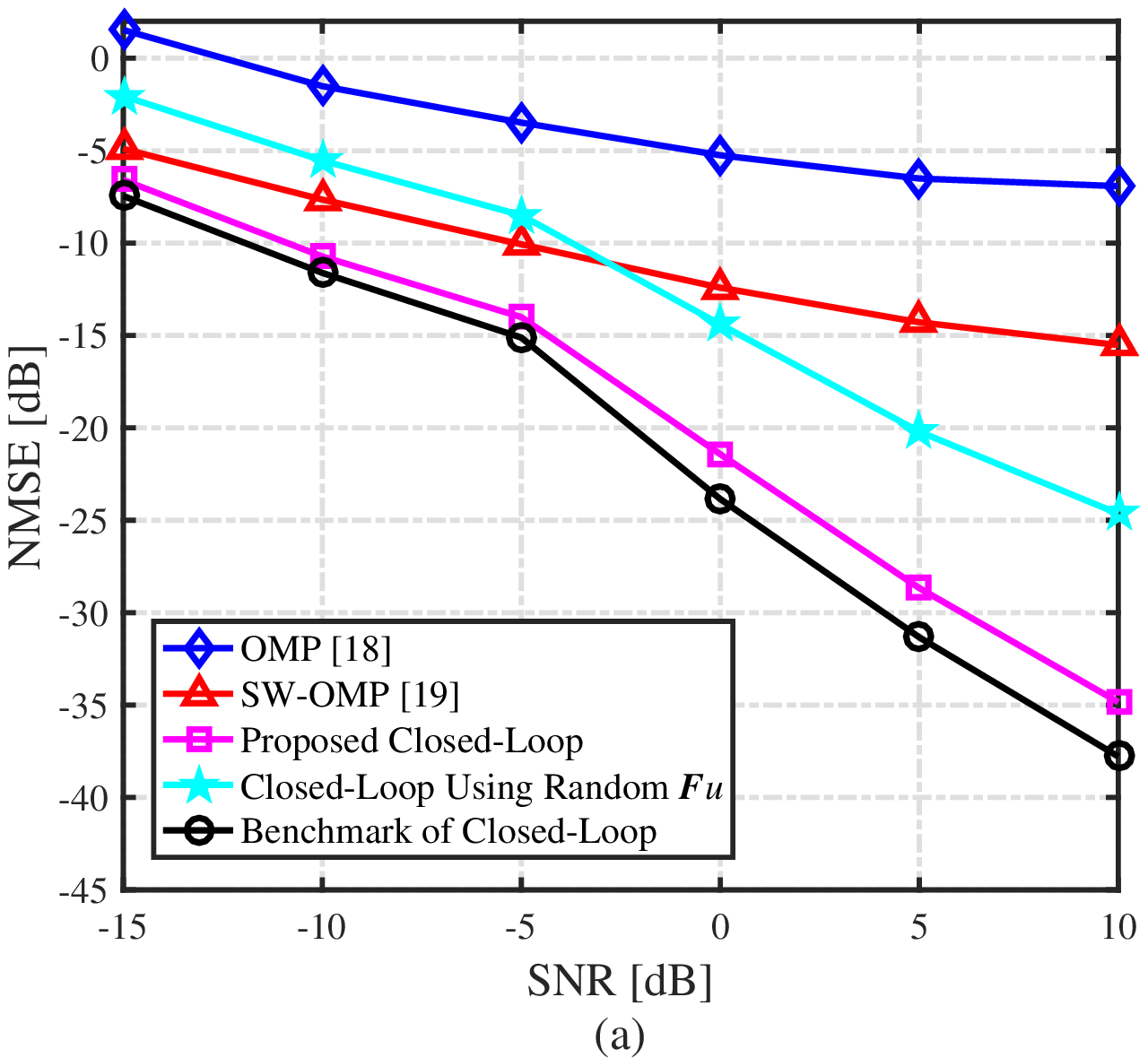}
\end{minipage}
}
\subfloat{
\label{FIG6(b)}
\begin{minipage}[t]{0.4\linewidth}
\centering
\includegraphics[width = 2.6in]{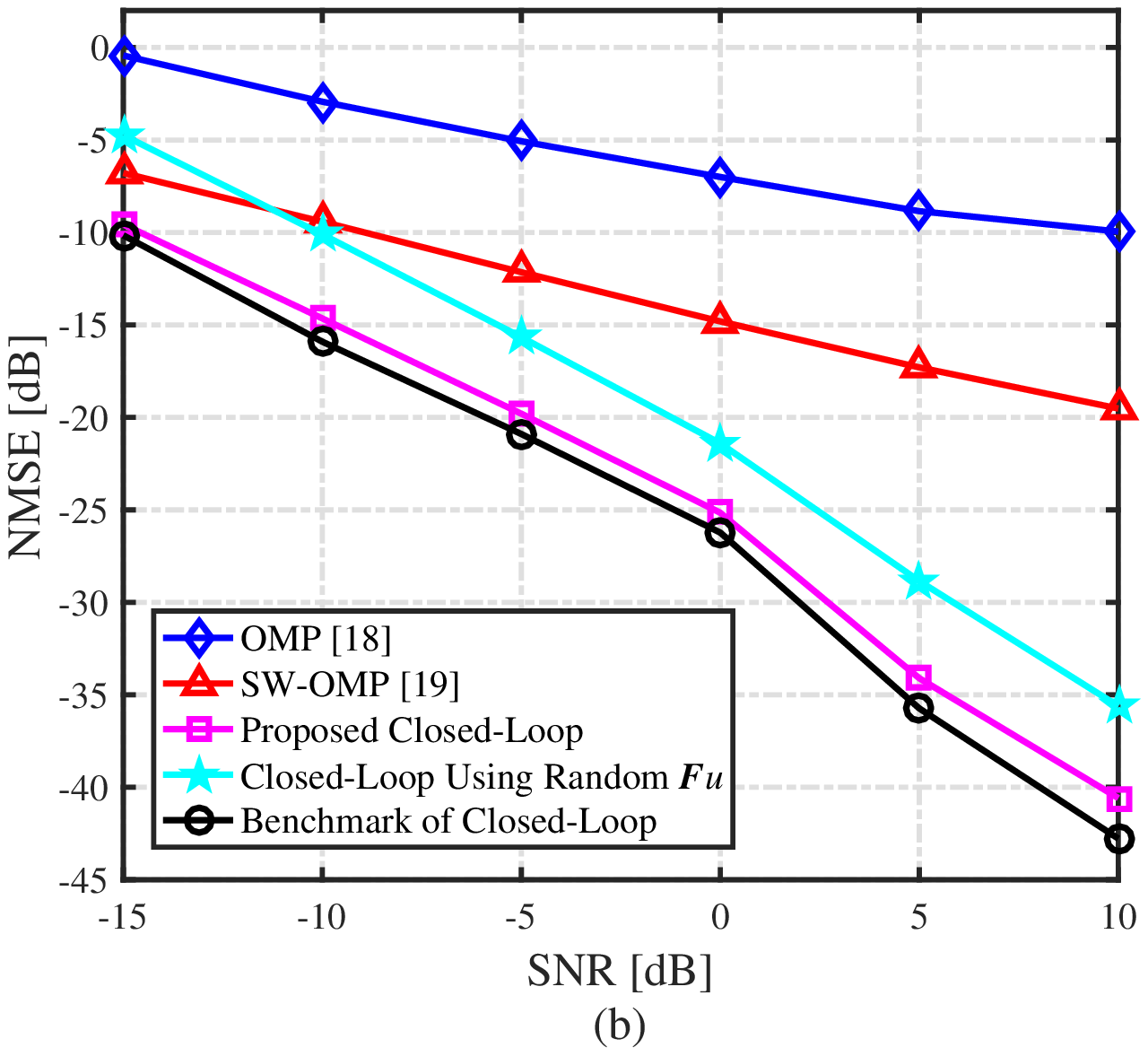}
\end{minipage}
}
\caption{NMSE performance comparison of different CE schemes versus SNRs with the same number of
 MPCs $L=4$: (a)~training overhead $T_{\rm CE}=88$; and (b) $T_{\rm CE}=176$.}
\label{FIG6}
\end{figure*}

\begin{figure*}[!tp]
\captionsetup{font={footnotesize}, name = {Fig.}, labelsep = period} 
\captionsetup[subfigure]{singlelinecheck = on, justification = raggedright, font={footnotesize}}
\centering
\subfloat{
\label{FIG7(a)}
\begin{minipage}[t]{0.4\linewidth}
\centering
\includegraphics[width = 2.6in]{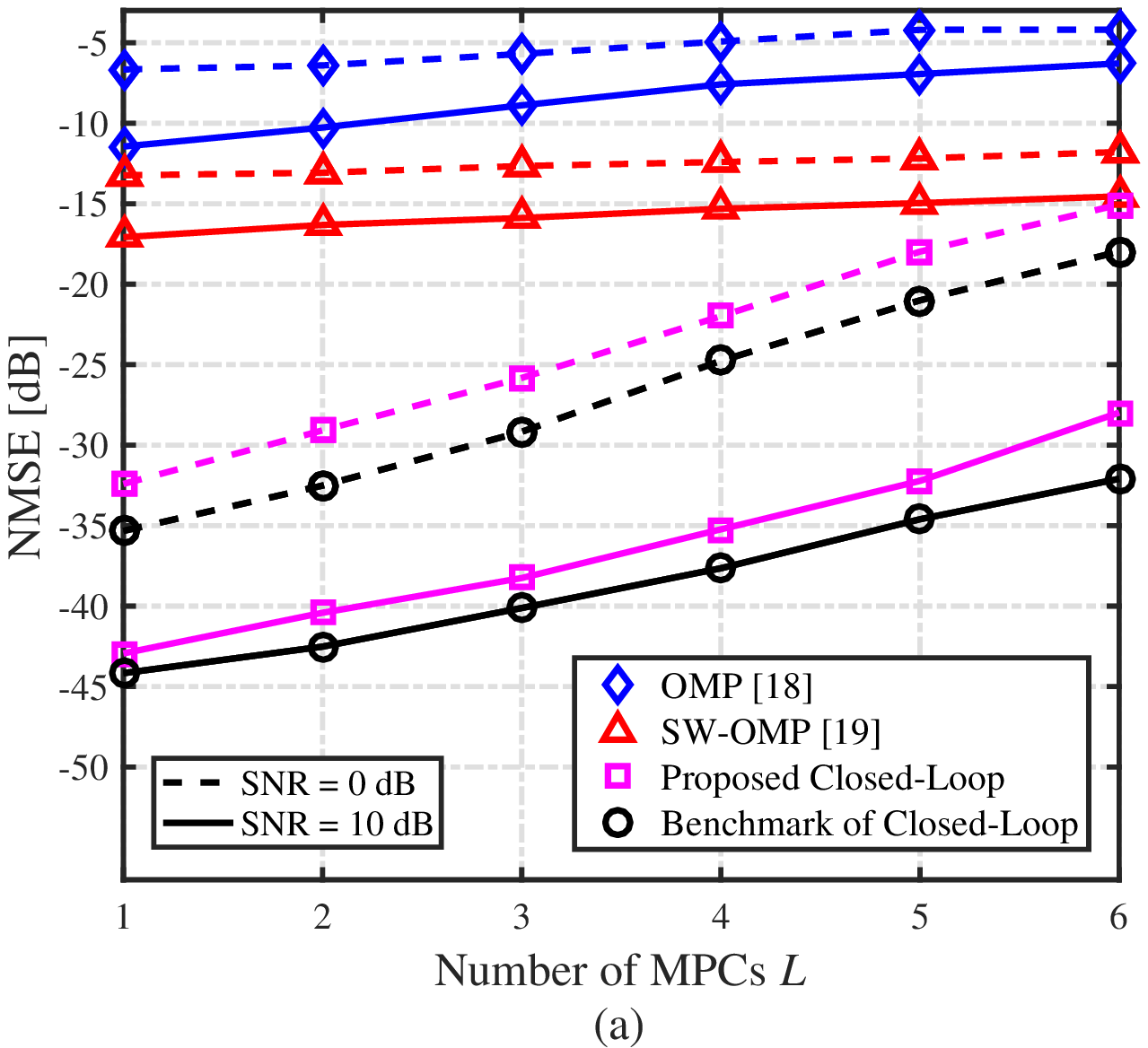}
\end{minipage}
}
\subfloat{
\label{FIG7(b)}
\begin{minipage}[t]{0.4\linewidth}
\centering
\includegraphics[width = 2.6in]{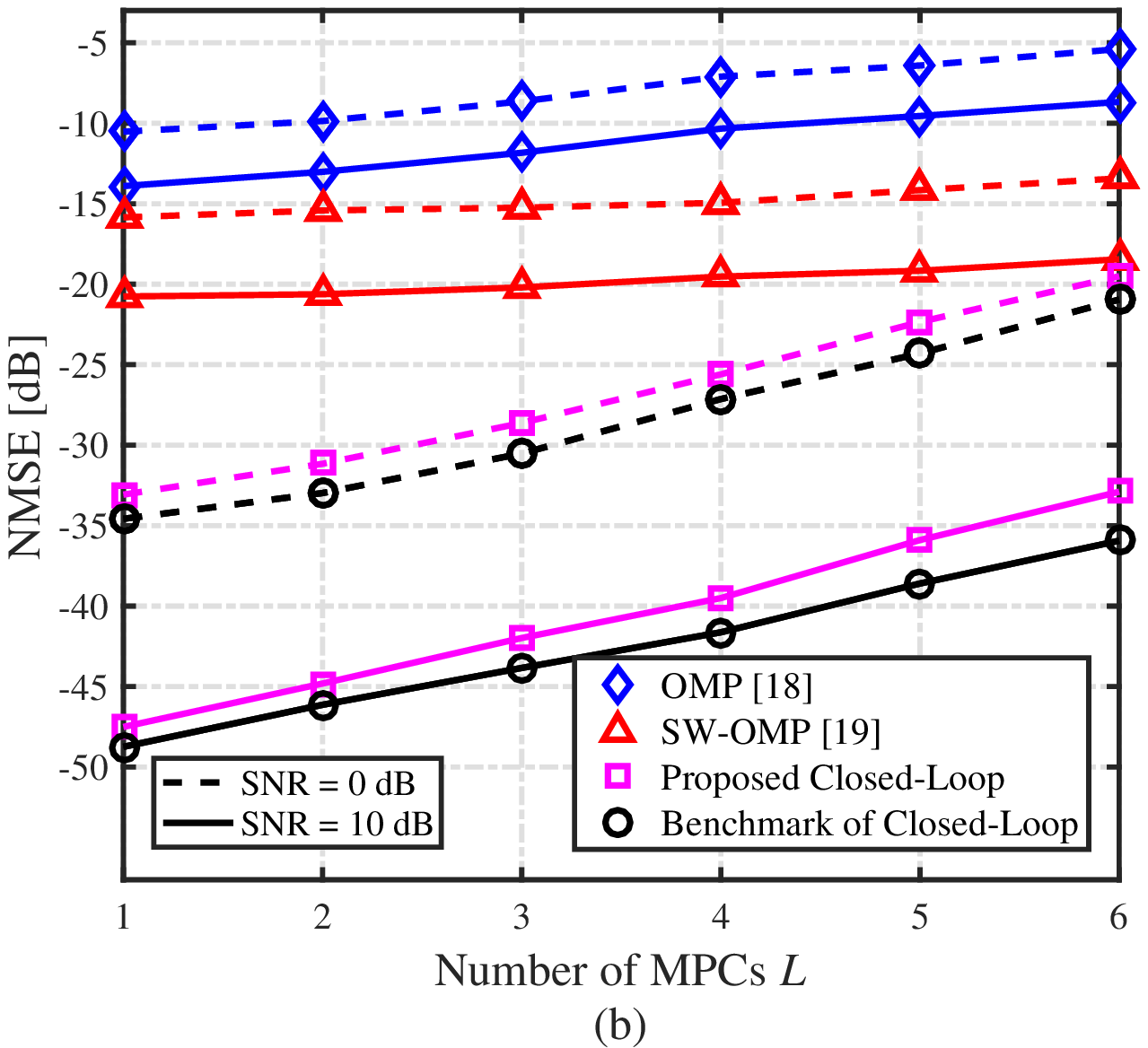}
\end{minipage}
}
\caption{NMSE performance comparison of different CE schemes versus the number of
 MPCs $L$ with $\text{SNR}=0$\,dB and 10\,dB: (a)~training overhead $T_{\rm CE}=88$;
 and (b)~$T_{\rm CE}=176$.}
\label{FIG7}
\end{figure*}

 Fig.~\ref{FIG5} compares the NMSE performance of the proposed closed-loop scheme with
 those of the OMP and SW-OMP based schemes for different SNRs, given the numbers of MPCs
 $L=3$ and $L=5$. For our closed-loop scheme, the numbers of OFDM symbols in each downlink
 time slot and uplink time slot are $N_{\rm o}^d \! =\! N_{\rm o}^u\! =\! 3$. Therefore,
 the training overhead of our closed-loop scheme is $T_{\rm CE}\! =\! N_d N_{\rm o}^d\!
 +\! N_u N_{\rm o}^u\! =\! 132$. In Fig.~\ref{FIG5}, the NMSE curve labeled as `Proposed
 Close-Loop' is our proposed closed-loop CE scheme, which also estimates the number of
 MPCs $L$, while the curve labeled with `Benchmark of Closed-Loop' is the closed-loop
 scheme given the perfect knowledge of $L$, which provides a lower bound NMSE. It can
 be seen that the CE accuracy achieved by our closed-loop scheme with no knowledge of
 $L$ is very close to this lower bound, which demonstrates the super-resolution accuracy
 of our solution Additionally, our closed-loop CE scheme adopting the random transmit
 precoding matrix $\bm{F}_u$ is also illustrated in Fig.~\ref{FIG5}, where it is observed
 to suffer from around 5\,dB and 3\,dB performance losses in the cases of $L=3$ and 5,
 respectively. This clearly demonstrates the effectiveness of the proposed multi-beam
 transmit precoding matrix design which fully exploits the estimated horizontal/vertical
 AoAs to optimize the received SNR for improving CE performance. Furthermore, the results
 of Fig.~\ref{FIG5} show that our proposed closed-loop CE scheme dramatically outperforms
 the two CS-based schemes, in terms of CE accuracy. In particular, the OMP and SW-OMP
 based schemes seem to suffer from the NMSE floor at high SNR. By adopting larger discrete
 angle-domain grids to achieve larger quantized CS dictionary, the performance of these
 CS-based schemes can be improved \cite{JSAC_OMP17,TWC_SW_OMP18} at the expense of
 significantly increased computational complexity, which becomes unaffordable for FD-MIMO
 systems with massive antenna array.

 Fig.~\ref{FIG6} compares the NMSE performance of different CE schemes against different
 SNRs, given two training overheads with the same number of MPCs $L=4$. $T_{\rm CE}\! =\! 88$
 in Fig.~\ref{FIG6(a)} and $T_{\rm CE}\! =\! 176$ in Fig.~\ref{FIG6(b)} correspond to
 choosing $N_{\rm o}^d\! =\! N_{\rm o}^u\! =\! 2$ and $N_{\rm o}^d\! =\! N_{\rm o}^u\! =\! 4$
 in our scheme, respectively. From Fig.~\ref{FIG6}, similar conclusions to those observed for
 Fig.~\ref{FIG5} can be obtained. In particular, it can be seen that our closed-loop CE scheme
 considerably outperforms the two CS-based schemes.

\begin{figure*}[!tp]
\captionsetup{font={footnotesize}, singlelinecheck = off, justification = raggedright, name = {Fig.}, labelsep = period} %
\captionsetup[subfigure]{singlelinecheck = on, justification = raggedright, font={footnotesize}}
\centering
\subfloat{
\label{FIG8(a)}
\begin{minipage}[t]{0.4\linewidth}
\centering
\includegraphics[width = 2.6in]{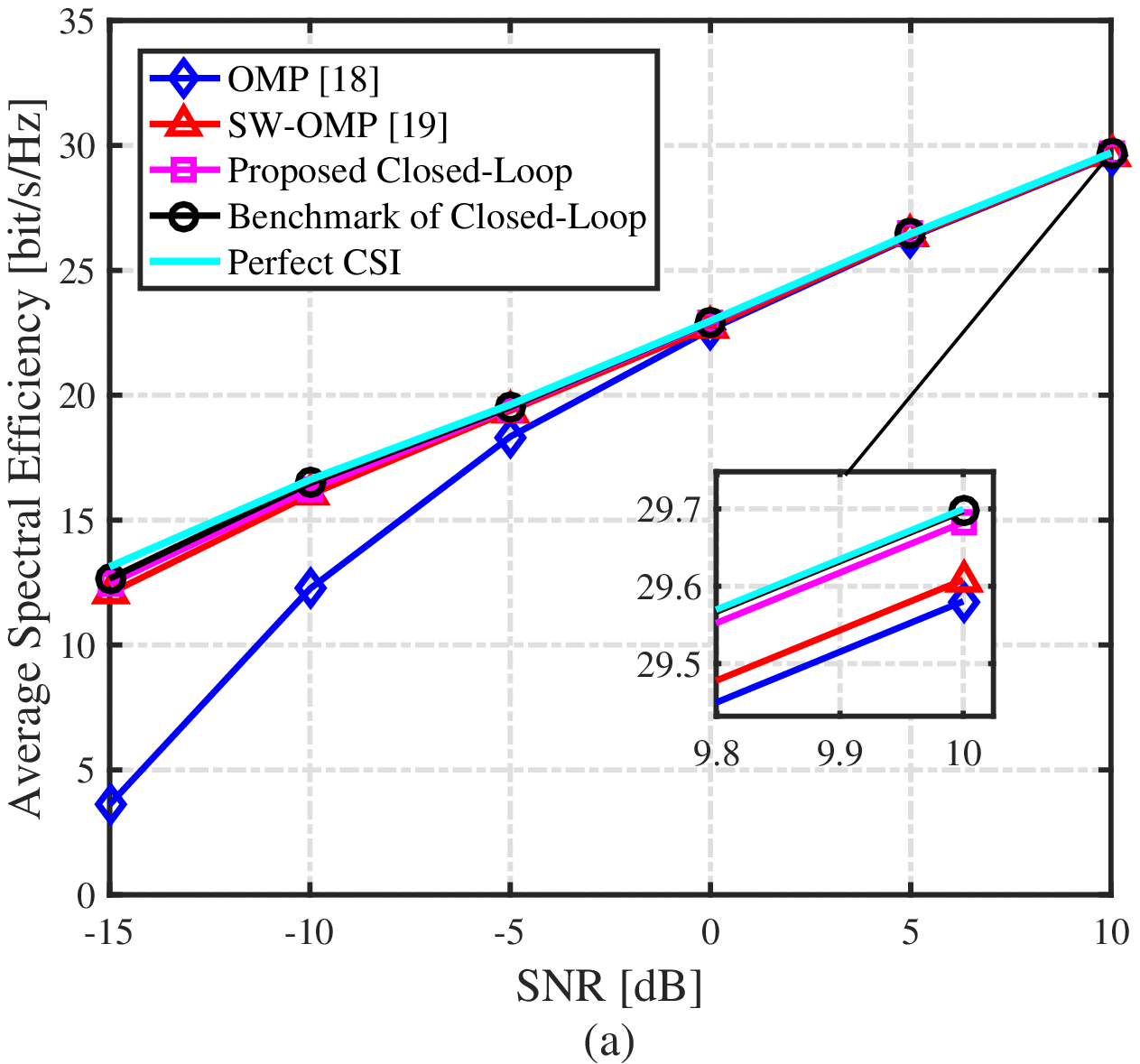}
\end{minipage}
}
\subfloat{
\label{FIG8(b)}
\begin{minipage}[t]{0.4\linewidth}
\centering
\includegraphics[width = 2.6in]{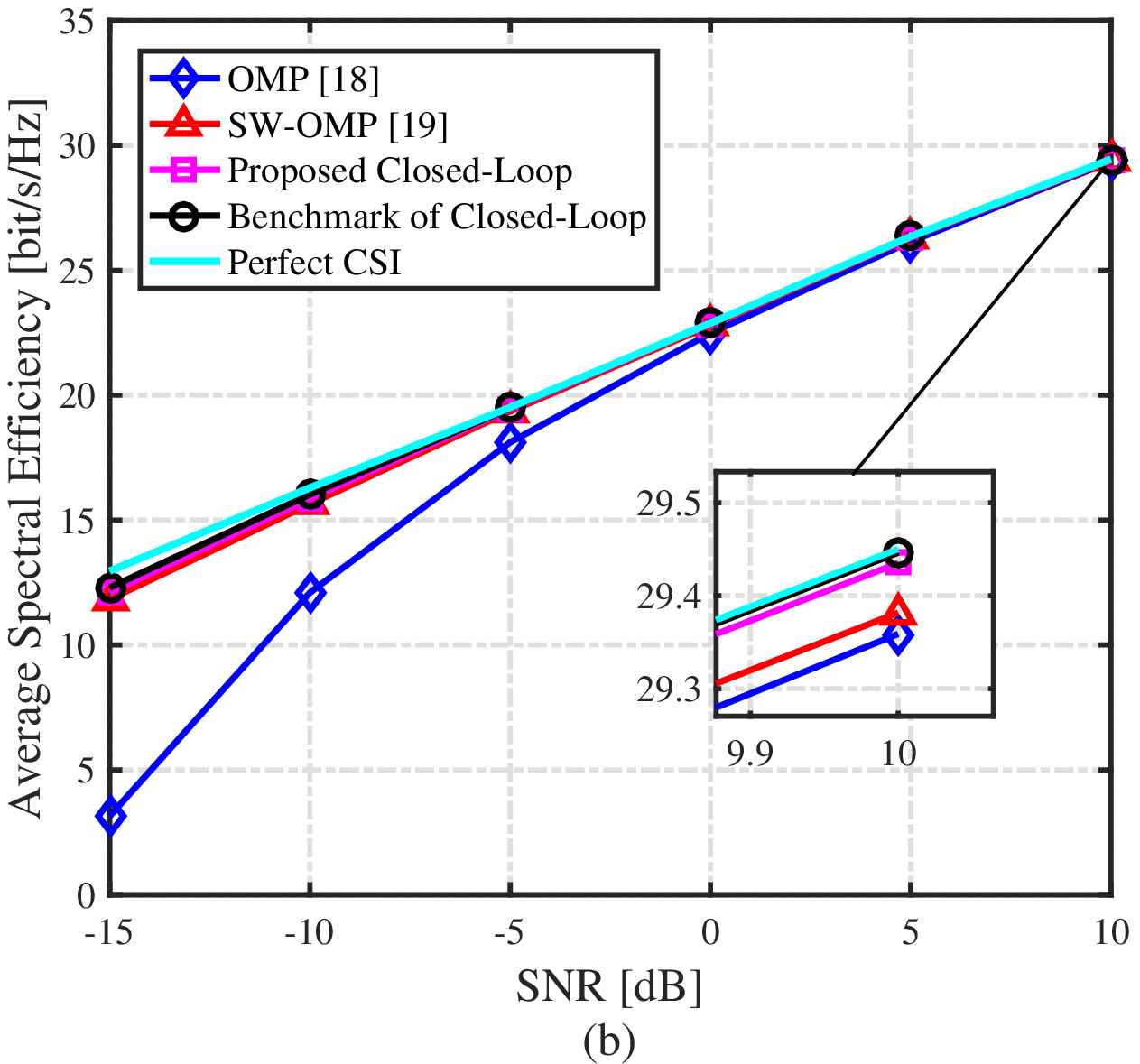}
\end{minipage}
}
\caption{ASE performance comparison of different CE schemes versus SNRs: (a)~the number of
 MPCs $L=3$; and (b)~$L=5$.}
\label{FIG8}
\end{figure*}

 Fig.~\ref{FIG7} compares the NMSE performance of different CE schemes versus the number
 of MPCs $L$ given two SNR values of 0\,dB and 10\,dB as well as two training overheads
 of $T_{\rm CE}=88$ and $T_{\rm CE}=176$. In simulations, we adopted the same parameter
 settings as in Fig.~\ref{FIG6} except the number of MPCs. From Fig.~\ref{FIG7}, the good
 performance of the proposed multi-beam transmit precoding matrix design and EVD-based
 approach for MPCs' number estimation is evident. Again, the proposed closed-loop scheme
 significantly outperforms two other CS-based schemes. Observe that the performance gain
 of our scheme over the two other schemes increases for sparser mmWave channels, i.e.,
 having smaller number of MPCs. Moreover, although the performance gap between the proposed
 solution and the CS-based methods is gradually reduced at low SNRs as the number of MPCs
 $L$ increases, the proposed scheme can achieve the considerable performance gain over
 the CS schemes at high SNRs. Hence, the proposed scheme is suitable for sparse mmWave
 channels, and more performance gain can be obtained for sparser channels.

\begin{table*}[!bp]
\renewcommand\arraystretch{1.5}
\captionsetup{labelsep = period} 
\caption{Computational Complexity of Proposed Closed-Loop Scheme}
\begin{center}
\label{TAB1}
\begin{tabular}{c|l}
\Xhline{1.2pt}
 Operation & \multicolumn{1}{c}{Complexity} \\
\Xhline{1pt}
 Step~1(b)\&5(b) & $\textsf{O}\left(N_d N_{\rm UD} N_{\rm UD}^{\rm RF}(1 + N_{\rm s}^d)
 + N_u N_{\rm BS} N_{\rm BS}^{\rm RF}(1 + N_{\rm s}^u)\right)$ \\
\hline
 Step 2 & $\textsf{O}\big( \left( N_{\rm UD}^{\rm sub}\right)^3 \!+\!
 \left( N_{\rm UD}^{\rm sub}\right)^2 N_{\rm o}^d N_P \big)$ \\
\hline
 \makecell{Step 3 \\ ($R = 2$)} & \makecell[l]{$\textsf{O}\bigg( \underbrace{M_{\rm sub}^{\rm UD} K N_{\rm o}^d
 G_1^d G_2^d}_{\rm 2-D\, SSP} \!\!+\!\! \underbrace{8 M_{\rm sub}^{\rm UD} K N_{\rm o}^d G_1^d
 G_2^d}_{\rm RVP} \!\!+\!\! \underbrace{\textstyle\frac{1}{4} M_{\rm sub}^{\rm UD} \widehat{L}^2}_{\rm SSA}
 \!\!+\!\! \underbrace{\textstyle\frac{1}{2}\left(\widehat{L}^3\!\!+\!\! 2\widehat{L}^2 M_{\rm sub}^{\rm UD}
 \!\!+\!\! 2(\widehat{L}\!\!+\!\! 1)( M_{\rm sub}^{\rm UD})^2\right)}_{\rm SIES} \!\!+\!\!
 \underbrace{\textstyle\frac{1}{4}\widehat{L}^3}_{\rm 2-D\, JD} \bigg)$} \\
\hline
 \makecell{Step 6 \\ ($R = 3$)} & \makecell[l]{$\textsf{O}\bigg( \underbrace{M_{\rm sub}^{\rm BS}
 N_{\rm o}^u G_1^u G_2^u G_3^u}_{\rm 3-D\, SSP} \!\!+\!\! \underbrace{8 M_{\rm sub}^{\rm BS} N_{\rm o}^u
 G_1^u G_2^u G_3^u}_{\rm RVP} \!\!+\!\! \underbrace{\textstyle\frac{1}{4} M_{\rm sub}^{\rm BS} \widehat{L}^2}_{\rm SSA}
 \!\!+\!\! \underbrace{\textstyle\frac{3}{4}\left( \widehat{L}^3 \!\!+\!\! 2\widehat{L}^2 M_{\rm sub}^{\rm BS}
 \!\!+\!\! 2(\widehat{L}\!\!+\!\! 1)( M_{\rm sub}^{\rm BS})^2\right)}_{\rm SIES} \!\!+\!\!
 \underbrace{\textstyle\frac{3}{4}\widehat{L}^4}_{\rm 3-D\, JD} \bigg)$} \\
\hline
 Step 7 & $\textsf{O}\big( \widehat{L}! (\widehat{L}^3 \!+\! 2\widehat{L}^2
 KN_{\rm BS}^{\rm sub} N_{\rm o}^u) \big)$\\
\hline
 Step 8 & $\textsf{O}\big( K \widehat{L} N_{\rm BS} N_{\rm UD} \big)$\\
\Xhline{1.2pt}
\end{tabular}
\end{center}
\end{table*}

 Next we consider the average spectral efficiency (ASE) performance metric \cite{Sun_PCA} defined as
\begin{equation}\label{Eq_ASE} 
\begin{split}
 \text{A}&\text{SE} = \frac{1}{K} \sum\nolimits_{k=0}^{K-1}\log_2 \det \big( \bm{I}_{N_{\rm s}}\\
 & + {\textstyle{1 \over {N_{\rm s}}}} \bm{R}_n^{-1}[k] \bm{W}_c^{\rm H}[k] \bm{H}[k] \bm{F}_p[k] \bm{F}_p^{\rm H}[k]
  \bm{H}^{\rm H}[k] \bm{W}_c[k] \big) ,
\end{split}
\end{equation}
 where $\bm{R}_n[k]\! =\! \sigma _n^2\bm{W}_c^{\rm H}[k]\bm{W}_c[k]$, $\bm{F}_p[k]\! =\!
 \bm{F}_{{\rm RF},p}\bm{F}_{{\rm BB},p}[k]$ and $\bm{W}_c[k]\! =\! \bm{W}_{{\rm RF},c}
 \bm{W}_{{\rm BB},c}[k]$ are the transmit precoding and receive combining matrices used
 during data transmission, respectively, while $N_{\rm s}$ is the number of transmit data
 stream. The principle component analysis (PCA)-based hybrid beamforming scheme proposed
 in \cite{Sun_PCA} is used to evaluate the ASE performance, where the CSI is based on
 the estimated channels. Besides, the spectral efficiency of the PCA-based hybrid
 beamforming scheme with the perfect CSI known both to the BS and UD is adopted as the
 performance upper bound. Here, the same simulation parameters used in Fig.~\ref{FIG5}
 are considered, and the number of transmit data streams used is $N_{\rm s}\! =\! 2$ in
 (\ref{Eq_ASE}). Fig.~\ref{FIG8} compares the ASE performance of different CE schemes
 against different SNRs. It can be observed from Fig.~\ref{FIG8} that the ASE performance
 using the CSI estimated by the proposed scheme closely matches to the performance upper
 bound obtained using the perfect CSI at both the BS and UD. It can also be seen that the
 ASE performance gain achieved by the proposed scheme over the two CS-based schemes is
 0.1\,[bit/s/Hz] at high SNR conditions. At low SNRs, this gain is clearly larger. Note
 that the ASE performance of the OMP based scheme is particularly poor when $\text{SNR}\!
 \le\! 0$\,dB.

\subsection{Computational Complexity Evaluation}\label{S6.2}

 The computational complexity analysis of our closed-loop CE scheme is detailed
 in Table~\ref{TAB1}, where the notation $\textsf{O}(N)$ stands for `on order of $N$'.
 The computational requirements of Step~1(a) and Step~5(a) are omitted, since they are
 much smaller, compared with the requirements of other steps. Clearly, Step~4 does not
 involve computation. It can be seen that the computational requirements are dominated
 by Step~6 (corresponding to Algorithm~\ref{ALG2} with $R\! =\! 3$) and Step~7. Also
 observe that the complexity of the CE scheme increases fast as $\widehat{L}$
 increases, since the computational complexity of Step~6 and Step~7 are proportional to
 $\widehat{L}^4$ and $\widehat{L} !$, respectively.

\begin{table*}[!tp]
\renewcommand\arraystretch{1.5}
\captionsetup{labelsep = period}
\centering
\caption{Computational Complexity of Two CS-Based CE Schemes}
\label{TAB2}
\begin{center}
\begin{tabular}{c|l!{\vrule width1pt}l}
\Xhline{1.2pt}
 Operation & \multicolumn{1}{c!{\vrule width1pt}}{\textbf{OMP-based Scheme} \cite{JSAC_OMP17}} &
  \multicolumn{1}{c}{\textbf{SW-OMP-based Scheme} \cite{TWC_SW_OMP18}} \\
\Xhline{1pt}
 Measurement matrix & $\textsf{O}\left( K T_{\rm CE} N_{\rm BS}^{\rm RF} N_{\rm BS} N_{\rm UD}
  G_{\rm BS} G_{\rm UD} \right)$ & $\textsf{O}\left( T_{\rm CE} N_{\rm BS}^{\rm RF} N_{\rm BS}
  N_{\rm UD} G_{\rm BS} G_{\rm UD} \right)$ \\
\hline
 Whitening & NA & $\textsf{O}\left( T_{\rm CE} N_{\rm BS}^{\rm RF} \left( ( T_{\rm CE}
  N_{\rm BS}^{\rm RF} )^2 + K + G_{\rm BS} G_{\rm UD} \right) \right)$ \\
\hline
 Correlation & $\textsf{O}\big(  T_{\rm CE} N_{\rm BS}^{\rm RF} G_{\rm BS} G_{\rm UD}
  \big(\sum\nolimits_{k=1}^K I_k\big) \big)$ & $\textsf{O}\left( T_{\rm CE}
  N_{\rm BS}^{\rm RF} G_{\rm BS} G_{\rm UD} K I \right)$ \\
\hline
 Project subspace & $\textsf{O}\big( \sum\nolimits_{k=1}^K
  \big( \frac{1}{4} I_k^2 \left( I_k + 1 \right)^2 + \frac{1}{3} I_k \left( I_k + 1 \right)
  \left( 2 I_k + 1 \right) T_{\rm CE} N_{\rm BS}^{\rm RF} \big)\big)$ & $\textsf{O}\big(
  \frac{1}{4} I^2 (I + 1)^2 + \frac{1}{3} K I (I + 1) (2I + 1) T_{\rm CE} N_{\rm BS}^{\rm RF} \big)$ \\
\hline
 Update residual & $\textsf{O}\big(  T_{\rm CE} N_{\rm BS}^{\rm RF} \big(\sum\nolimits_{k=1}^K
  \frac{I_k}{2}\left( I_k + 1 \right)\big) \big)$ & $\textsf{O}\big( T_{\rm CE} N_{\rm BS}^{\rm RF}
  K \frac{I}{2} (I + 1) \big)$ \\
\hline
 Compute MSE & $\textsf{O}\big( T_{\rm CE} N_{\rm BS}^{\rm RF} \big(\sum\nolimits_{k= }^K I_k\big)
  \big)$ & $\textsf{O}\left( T_{\rm CE} N_{\rm BS}^{\rm RF} K^2 I \right)$ \\
\hline
 Reestablishment & $\textsf{O}\big( N_{\rm BS} N_{\rm UD} \big(\sum\nolimits_{k=1}^K I_k\big)
  \big)$ & $\textsf{O}\left( N_{\rm BS} N_{\rm UD} K I \right)$ \\
\Xhline{1.2pt}
\end{tabular}
\end{center}
\end{table*}

\begin{figure*}[!tp]
\captionsetup{font={footnotesize}, name = {Fig.}, labelsep = period} 
\captionsetup[subfigure]{singlelinecheck = on, justification = raggedright, font={footnotesize}}
\centering
\subfloat{
\label{FIG9(a)}
\begin{minipage}[t]{0.4\linewidth}
\centering
\includegraphics[width = 2.6in]{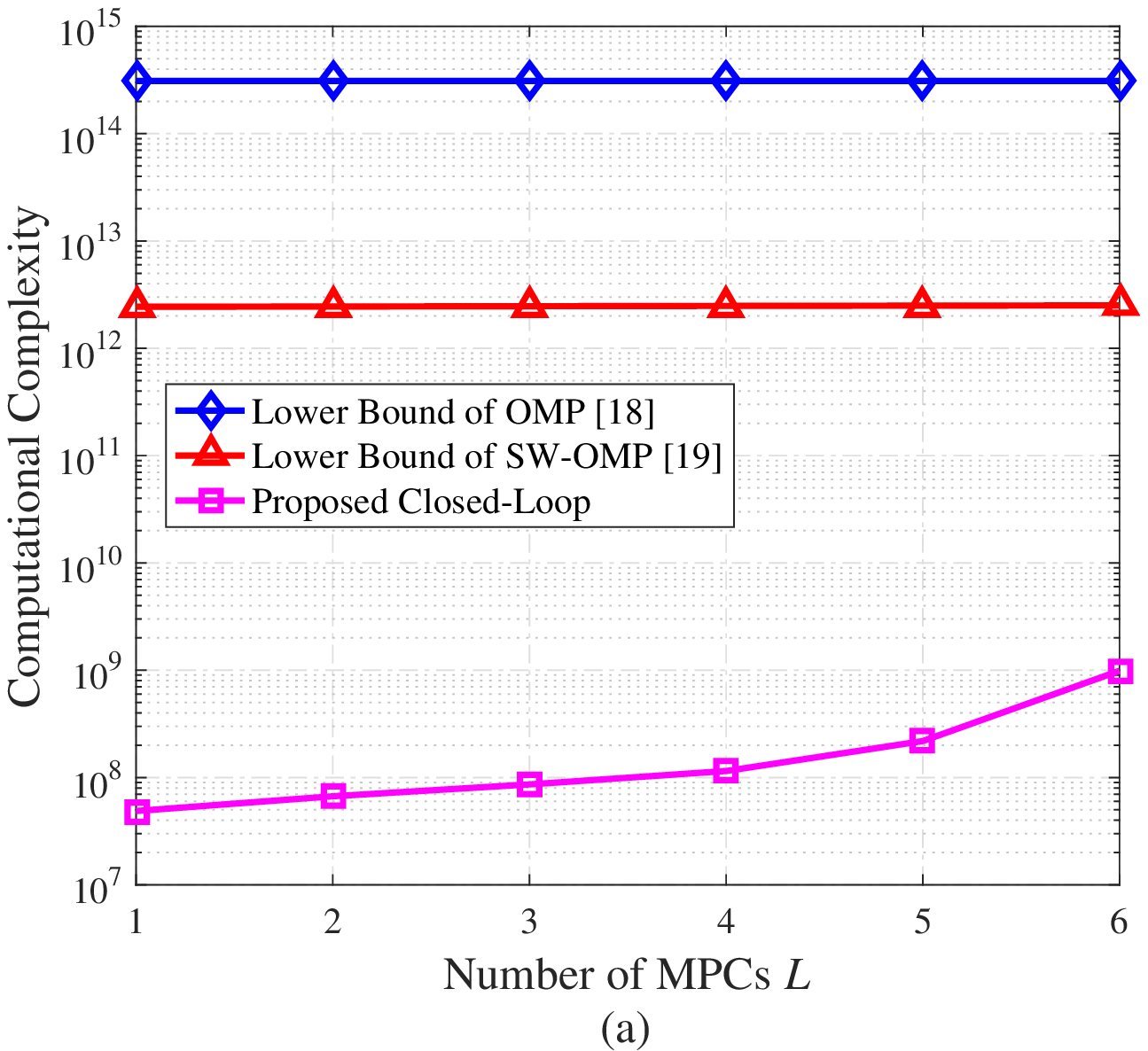}
\end{minipage}
}
\subfloat{
\label{FIG9(b)}
\begin{minipage}[t]{0.4\linewidth}
\centering
\includegraphics[width = 2.655in]{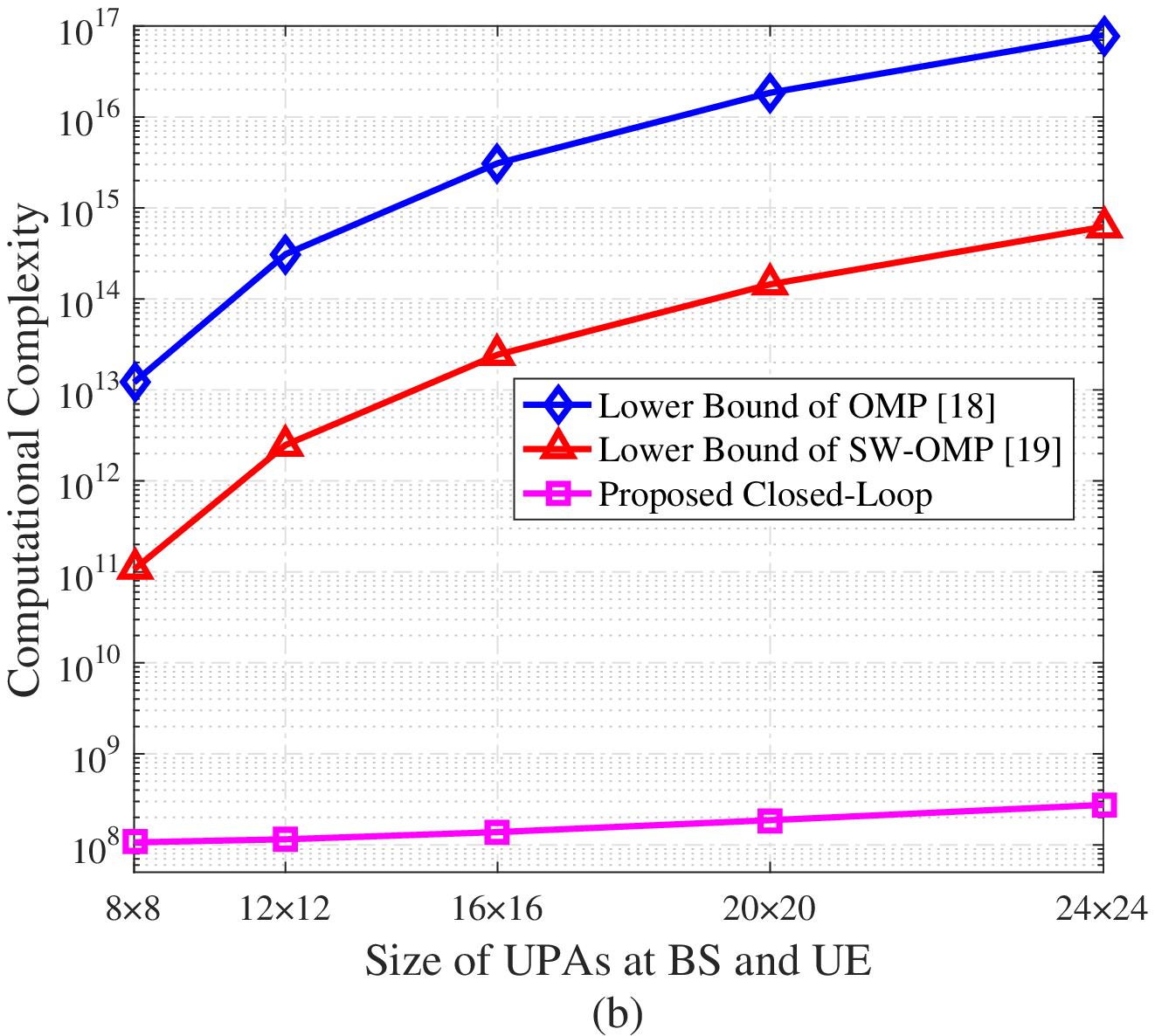}
\end{minipage}
}
\caption{Computational complexity comparison of different CE schemes given the training
 overhead $T_{\rm CE}=88$: (a)~sizes of UPAs at BS and UD are both $12 \times 12$;
 and (b)~number of MPCs $L = 4$.}
\label{FIG9}
\end{figure*}

 The computational complexity of the two CS-based CE schemes is given in Table \ref{TAB2}
 for comparison, where the numbers of iterations for the OMP algorithm at the $k$th subcarrier
 and the SW-OMP algorithm are denoted by $I_k$ and $I$, respectively. Note that the values
 of $I_k$ are different for different subcarriers. It can be seen that the computational
 complexity of these two CS-based schemes increase fast as the quantized grids
 $G_{\rm BS}$ and $G_{\rm UD}$ increase. Also the complexity of the OMP scheme is around
 $K$ times of the SW-OMP scheme, because the $K$ subchannels at $K$ subcarriers are
 independently estimated in the OMP scheme but they are jointly estimated in the SW-OMP scheme.
 Due to the power leakage caused by the mismatch between the discrete CS angle-domain
 dictionary and continuously distributed AoAs/AoDs of channels, the number of effective MPCs
 represented in the redundant CS dictionary are usually greater than $L$. Hence, the value of
 $I_k$ in the OMP scheme and the value of $I$ in the SW-OMP scheme are not fixed and they
 are usually greater than $L$. Therefore, we can use $I\! =\! I_k \! =\! L$ to provide the
 lower bounds of the computational complexity for the two CS-based schemes.

 Fig.~\ref{FIG9} compares the computational complexity of our closed-loop CE scheme with
 those of the two CS-based schemes given the training overhead $T_{\rm CE}=88$ corresponding
 to $N_{\rm o}^d\! =\! N_{\rm o}^u\! =\! 2$ in our scheme. From Fig.~\ref{FIG9(a)}, we
 observe that the computational complexity of the proposed CE solution increases slightly
 as the number of MPCs increase. Most strikingly, however, given the size of UPA as
 $12\times 12$, the complexity of our solution is at least 3 orders of magnitude lower
 than the SW-OMP scheme and at least 5 orders of magnitude lower than the OMP scheme.
 The results of Fig.~\ref{FIG9(b)} indicate that given $L\! =\! 4$, the complexity of
 our solution is almost immune to the size of UPA at the BS and UD. By contrast, the
 complexity of the two CS-based schemes increase considerably as the number of antennas
 increases. Again, the complexity of our solution is several orders of magnitude lower
 than the other two schemes. It should also be reiterated that to mitigate the power
 leakage, the CS-based schemes adopt the high-dimensional redundant dictionary, which
 results in unaffordable storage space requirements when the number of antennas is large.
 Clearly, for FD-MIMO systems with massive number of antennas, the proposed closed-loop
 scheme offers considerable advantage over the CS-based schemes, in terms of both
 computational complexity and storage requirements.

\begin{figure*}[!tp]
\captionsetup{font = normalsize, labelsep = period} 
\caption*{TABLE III. Comparison of Advantages and Disadvantages of Different CE Schemes} \label{TableIII}
\begin{center}
\includegraphics[width = 2.0\columnwidth,keepaspectratio]{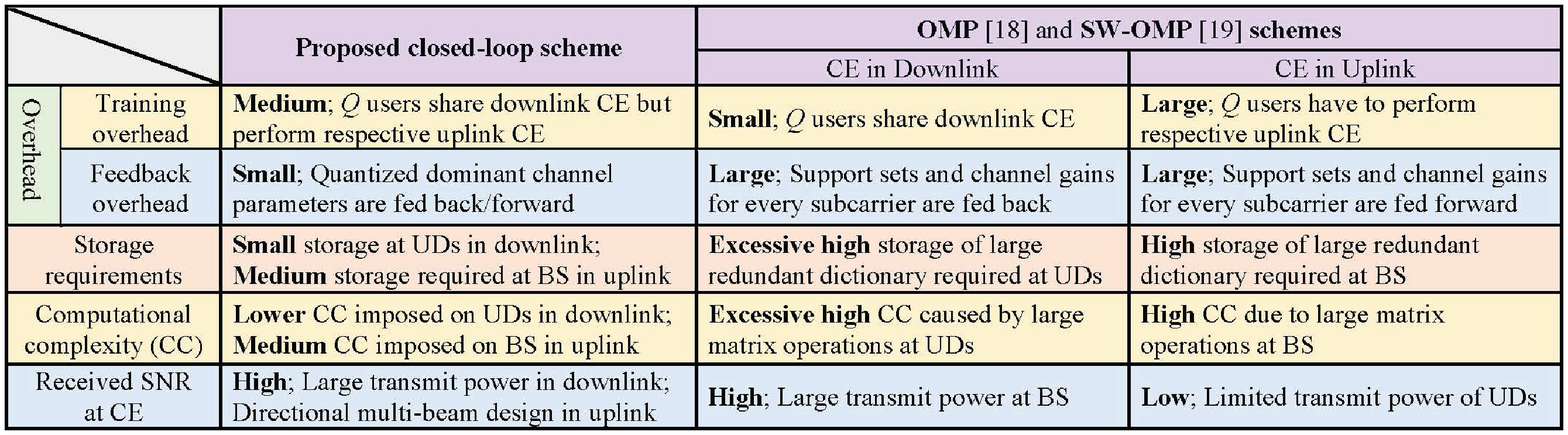}
\end{center}
\end{figure*}

 The advantages and disadvantages of our proposed solution and two other CS-based CE schemes
 are given in Table III, where the training/feedback overhead, storage requirements,
 computational complexity and received SNR at CE are compared.

\section{Conclusions}\label{S7}

 We have proposed a closed-loop sparse CE scheme for multi-user wideband mmWave FD-MIMO
 systems with hybrid beamforming. By exploiting the sparsity of mmWave channels in
 both angle and delay domains and by visualizing high-dimensional hybrid arrays as
 low-dimensional digital arrays, the proposed scheme is capable of obtaining the
 super-resolution estimates of horizontal/vertical AoDs/AoAs and delays based on
 the MDU-ESPRIT algorithm. Specifically, at the downlink CE stage, we design the
 common random transmit precoding matrix at the BS and the receive combining matrix
 at each UD to estimate the horizontal/vertical AoAs of sparse MPCs. At the uplink
 CE stage, based on the designed receive combining matrix at the BS, we estimate
 horizontal/vertical AoDs and delays. Furthermore, the AoAs estimated at each UD are
 utilized to design the multi-beam transmit precoding matrix for further enhancing
 CE performance. We also propose an ML approach at the BS to pair the channel
 parameters acquired at the two stages and to optimally estimate the path gains.
 Simulation results have demonstrated that the proposed closed-loop CE scheme offers
 considerable advantages over state-of-the-art CS-based CE schemes, in terms of
 providing significantly more accurate CSI estimate while imposing dramatically lower
 computational complexity and storage requirements.

\appendix

 Sampling the delay-domain continuous $\bm{H}_q(\tau )$ in (\ref{delay_domain_channel})
 with the sampling period $T_s$ yields
\begin{equation}\label{sampling_channel}
\begin{split}
 \bm{H}_q(n T_s) &= \beta_q \sum\limits_{l=1}^{L_q} \bm{H}_{q,l} p\left(\tau\! -\! \tau _{q,l}\right)
  \circledast \sum\limits_{n=-\infty}^{\infty} \delta\left(\tau\! -\! n T_s\right)\\[-1.5mm]
  &= \beta_q \sum\limits_{n=-\infty}^{\infty} \sum\limits_{l=1}^{L_q} \bm{H}_{q,l} p\left( n T_s -\tau_{q,l}\right) ,
\end{split}
\end{equation}
 where $\circledast$ and $\delta (\cdot )$ represent the linear convolution operation
 and Dirac delta function, respectively. The Fourier transform of $\bm{H}_q(n T_s)$ is
 then given by
\begin{equation}\label{Frequency_domain_channel}
 \bm{H}_q(f)\! =\! \frac{\beta_q}{T_s} \sum\limits_{l=1}^{L_q} \sum\limits_{n=-\infty}^{\infty}
  \bm{H}_{q,l} P(f) e^{-\textsf{j}2\pi f \tau_{q,l}} \delta\left( f\! -\! n f_s\right) ,
\end{equation}
 where $P(f)$ is the Fourier transform of $p(\tau )$. Obviously, $\bm{H}_q(f)$ exhibits
 periodicity with period $f_s$. Thus, $\bm{H}_q(f)$ within a period of
 $f\in \left[ -f_s/2, ~ f_s/2 \right]$ can be expressed as
\begin{equation}\label{Frequency_domain_channel_f}
 \bm{H}_q(f)\! =\! \frac{\beta_q}{T_s} \sum\limits_{l=1}^{L_q} P(f) \bm{H}_{q,l}
  e^{-\textsf{j}2\pi f \tau_{q,l}}\! \approx\! \frac{\beta_q}{T_s} \sum\limits_{l=1}^{L_q}
  C  \bm{H}_{q,l} e^{-\textsf{j}2\pi f \tau_{q,l}} .
\end{equation}
 The approximation in (\ref{Frequency_domain_channel_f}) is valid because the PSF
 $p(\tau )$ is designed to realize the ideal passband filter characteristics of
 $P(f)=C$ for $f\in\left[ -f_s/2, ~ f_s/2\right]$ and $P(f)\approx 0$ for
 $f\notin \left[ -f_s/2, ~ f_s/2\right]$. For convenience, we consider $C=T_s$.
 Therefore, the frequency-domain channel matrix $\bm{H}_q[k]$ at the $k$th subcarrier,
 where $0\le k\le K-1$, can be written as
\begin{equation}\label{frequency_domain_channel_k}
 \bm{H}_q[k] = \bm{H}_q\bigg(\frac{k f_s}{K}\bigg) = \beta_q \sum\limits_{l=1}^{L_q}
  \bm{H}_{q,l} e^{-\textsf{j}\frac{2\pi k f_s \tau_{q,l}}{K}} .
\end{equation}

\ifCLASSOPTIONcaptionsoff
  \newpage
\fi

\end{document}